\documentclass[aps,prd,showpacs,nofootinbib,preprintnumbers,twocolumn]{revtex4}
\usepackage{bm}
\usepackage{latexsym}
\usepackage{natbib}
\usepackage{url}
\usepackage{dcolumn}
\usepackage{color}
\usepackage{amsfonts,amssymb,amsmath}
\usepackage{graphicx,epsfig}
\usepackage{psfrag}
\usepackage{subfigure}
\usepackage{natbib}
\usepackage{hyperref}
\usepackage{comment}

\begin{document}
\title{Perturbativity, vacuum stability and inflation in the light of 750 GeV diphoton excess}

\author{Mansi Dhuria}
\email{mansi@prl.res.in}
\affiliation{Theoretical Physics Division, Physical Research Laboratory, \\Navrangpura, Ahmedabad 380 009, India}
\author{Gaurav Goswami}
\email{gaurav.goswami@ahduni.edu.in}
\affiliation{Institute of Engineering and Technology, Ahmedabad University, \\Navrangpura, Ahmedabad 380 009, India}

\date{\today}
\begin{abstract}
The recent observation of the 750 GeV diphoton excess at 13 TeV LHC has motivated many scenarios of physics 
beyond the Standard Model. 
In this work, we begin by showing that many models which explain the observed excess tend to get strongly coupled well below the Planck scale.
We then study a simple scenario involving colored vector-like fermions with exotic charges, which is expected to stay weakly coupled till the Planck scale. We find the conditions under which this happens, 
derive the Renormalization Group equations for such models and solve them to show that perturbativity till Planck scale 
can be maintained for a very reasonable choice of parameters.
Finally, we discuss issues related to vacuum stability and the possibility of inflation in the scenarios we study. \end{abstract}
\maketitle

\section{Introduction}
Though the existence of physics beyond the Standard Model (SM) of elementary particle physics is a widespread belief, new 
physics is yet to show up at accessible energies at the colliders.
However, recently, both the CMS and ATLAS collaboration have recently reported an excess in the diphoton channel at around 750 GeV from the data collected  at  the LHC run 2 at energy $\sqrt{s}= 13$ TeV \cite{CMS:2015dxe, ATLAS}. The ATLAS has analysed 3.2 fb$^{-1}$ of the data collected at $13$ TeV,  reporting  an  excess of about 14 $\gamma \gamma$ events  peaked at $M = 747$ GeV. With this, the best fit value of cross-section suggests high width $\Gamma/M \sim 0.06$ with a local significance of 3.9$\sigma$. Similarly, the CMS experiment has analysed  2.6 fb$^{-1}$  of the data collected at $ 13$ TeV  and reported an excess of  about 10 $\gamma \gamma$ events peaked at $M = 760$ GeV. The best fit value of the cross-section in this case suggests a very narrow width  with a local significance of 2.6$\sigma $. However, the CMS best-fit width matches with the value provided by ATLAS by considering the local significance of the excess  at CMS upto 2.0$\sigma$.  The CMS has also presented the compatibility of the excess at both $ 8 $ TeV \cite{CMS:2015cwa} and  $ 13 $ TeV \cite{CMS:2015dxe},  suggesting a production cross-section times branching ratio into photons to be $4.47 \pm 1.86$ fb at CMS \cite{CMS:2015dxe}. In case of ATLAS experiment, the compatibility of excess at both $ 8 $ TeV \cite{Aad:2015mna} and  $ 13 $ TeV \cite{ATLAS} suggests production cross-section times branching ratio into photons to be $10.6 \pm 2.9$ fb \cite{DiChiara:2015vdm}. 

Although the observation requires further analysis and much more data to ensure that the excess signal is not just a statistical fluctuation, it is still tempting to study phenomenological implications of this result 
\cite{Mambrini:2015wyu,Harigaya:2015ezk,Nakai:2015ptz,Buttazzo:2015txu,Franceschini:2015kwy,DiChiara:2015vdm,Angelescu:2015uiz,Pilaftsis:2015ycr,Bellazzini:2015nxw,Knapen:2015dap,Ellis:2015oso,Gupta:2015zzs,Molinaro:2015cwg,Higaki:2015jag,McDermott:2015sck,Low:2015qep,Petersson:2015mkr,Cao:2015pto,Matsuzaki:2015che,Dutta:2015wqh,Kobakhidze:2015ldh,Cox:2015ckc,Ahmed:2015uqt,Agrawal:2015dbf,Becirevic:2015fmu,No:2015bsn,Demidov:2015zqn,Chao:2015ttq,Fichet:2015vvy,Curtin:2015jcv,Bian:2015kjt,Chakrabortty:2015hff,Csaki:2015vek,Falkowski:2015swt,Aloni:2015mxa,Bai:2015nbs,Benbrik:2015fyz,Kim:2015ron,Gabrielli:2015dhk,Alves:2015jgx,Megias:2015ory,Carpenter:2015ucu,Bernon:2015abk,Ringwald:2015dsf,Arun:2015ubr,Han:2015cty,Chang:2015bzc,Han:2015dlp,Luo:2015yio,Chang:2015sdy,Bardhan:2015hcr,Feng:2015wil,Barducci:2015gtd,Chao:2015nsm,Chakraborty:2015jvs,Ding:2015rxx,Hatanaka:2015qjo,Antipin:2015kgh,Wang:2015kuj,Cao:2015twy,Huang:2015evq,Bi:2015uqd,Kim:2015ksf,Berthier:2015vbb,Cline:2015msi,Chala:2015cev,Kulkarni:2015gzu,Dev:2015isx,Boucenna:2015pav,deBlas:2015hlv,Murphy:2015kag,Hernandez:2015ywg,Dey:2015bur,Pelaggi:2015knk,Belyaev:2015hgo,Huang:2015rkj,Cao:2015xjz,Gu:2015lxj,Moretti:2015pbj,Patel:2015ulo,Badziak:2015zez,Chakraborty:2015gyj,Altmannshofer:2015xfo,Cvetic:2015vit,Allanach:2015ixl,Davoudiasl:2015cuo,Das:2015enc,Cheung:2015cug,Liu:2015yec,Zhang:2015uuo,Casas:2015blx,Hall:2015xds}. 
The observed resonance implies the existence of a spin-0 or spin-2 state since 
a spin-1 state can not decay into two photons due to the Landau-Yang theorem \cite{Landau:1948kw,Yang:1950rg}.  

We assume that the observed excess is due to SM singlet spin-zero particle. Since the singlet scalar does not have any direct interaction with photons, it has to decay into two photons  through a loop of electrically 
charged particles. There exist a vast literature on diphoton excess in which the particles running in the loop are 
assumed to be vector-like fermions \cite{Franceschini:2015kwy,Pilaftsis:2015ycr,Ellis:2015oso}.
Thus, $750$ GeV diphoton excess motivates a BSM scenario involving the following particles: all SM particles, 
a SM singlet scalar and charged vector-like fermions.
Moreover, the existence of vector-like fermions can in general be motivated by many theories of Beyond Standard Model (BSM) physics, such as string-theoretic models \cite{Dijkstra:2004cc,Lebedev:2006kn}, little Higgs models \cite{Han:2003wu}, and composite Higgs models  \cite{Contino:2006qr} etc.

It is noteworthy that the search for the vector-like fermions at LHC run 2 has pushed the limit of vector-like quarks upto around TeV \cite{CMS:2015alb,Khachatryan:2015oba, Aad:2016qpo}. Motivated by this, we 
consider the mass of these new fermions to be 1.2 TeV.
We show that the compatibility of the observed diphoton signal rate generically gives a very high value of 
Yukawa coupling of real singlet scalar with vector-like fermion, thus making the theory strongly coupled.
We then show that the issue can be resolved either by considering additional copies of vector-like fermions or/and by choosing colored vector-like fermions to have exotic charges (i.e. $Q > 2/3$).

It is well known that if SM is assumed to hold good till arbitrarily high energies, the Higgs potential turns negative at high field values (above $10^{10}$ GeV) suggesting the existence of new physics beyond the SM. 
It is then very interesting to study the impact of the new particles needed to explain the 
diphoton excess on the electroweak vacuum stability of the SM. 
We present a detailed study of the effect of these extra particles on the scalar potential of the theory. 
E.g. we find the conditions under which all couplings can be kept small and positive till Planck scale.
We then look into the possibility of the new scalar being responsible for inflation, similar to the SM Higgs.
In order to study the consequences of this model at high energies in detail, we have found the 
Renormalization Group evolution equations for scenarios involving extra vector like fermions with exotic charges.

The outline of the article is as follows: In section {\bf 2}, we begin by finding
the generic conditions under which the theory remains perturbative and weakly coupled if the diphoton excess  is explained by the decay of a SM singlet scalar by considering heavy colored vector-like fermion(s) in the loop. 
In section {\bf 3},  we present two simple models with exotic charge fermions which could not only explain the diphoton excess 
but also maintain perturbativity upto Planck scale. To achieve this, we ensure, among other things, 
that the mixing of the newly added scalar with SM Higgs is small and
that none of the couplings, such as the gauge coupling $g_1$, become too large below Planck scale.
In section {\bf 4}, we show that the proposed models (a) stay weakly coupled till Planck scale, (b) do not suffer from the vacuum stability problem, and,  
(c) are such that both SM Higgs and the 750 GeV scalar could act as the inflaton.
For this, we solve the renormalization group (RG) evolution equations for all couplings involved in the two models.
Moreover, we find that only one of the two models is consistent with the requirement of not leaving any dangerous relics in the Universe.
Finally, in section {\bf 5}, we summarize our results.  
In Appendix {\bf A} we provide RG equations generalized to the scenarios which we study in this paper.

\section{The various constraints}

We interpret the diphoton signal as being due to a 750 GeV scalar (as opposed to a pseudo scalar) $S$ produced by gluon fusion and 
then decaying into two photons. 
The cross-section for this process is given by \cite{Gabrielli:2015dhk},
\begin{equation} \label{eq:sigma-gluon-photon}
\sigma (gg \rightarrow S \rightarrow \gamma \gamma) = \frac{\pi^2}{8 m_s^3}~ {\cal I}_{\rm pdf}~ \Gamma (S \rightarrow \gamma \gamma) \; ,
\end{equation}
where, $m_s$ is the mass of the scalar and ${\cal I}_{\rm pdf}$ is the dimensionless parton distribution function integral evaluated at $\sqrt{s} = 13$ TeV. 
Let $g_{s \gamma}$ be the coefficient of the dimension five operator 
$\phi F^{\mu \nu} F_{\mu \nu}$ in the low energy effective field theory.
Then, the decay width $\Gamma(S \rightarrow \gamma \gamma)$ is given by 
$\Gamma(S \rightarrow \gamma \gamma) = g_{s \gamma}^2 m_s^3/(8 \pi)$. Using this and 
${\cal I}_{\rm pdf} \approx 5.8$ (by assuming $m_S=750$ GeV, see \cite{Gabrielli:2015dhk}), we can find $g_{s \gamma}$ in terms of the total cross-section 
$\sigma (pp \rightarrow S \rightarrow \gamma \gamma)$ as
\begin{equation} \label{eq:gsigma}
\frac{g_{s \gamma}}{({\rm TeV})^{-1}} = 3 \times 10^{-3} \left( \frac{\sigma}{\rm f b} \right)^{1/2} \; .
\end{equation}
The cross-section  $\sigma (pp \rightarrow S \rightarrow \gamma \gamma)$ is inferred by ATLAS and CMS experiments to be in 
the range $[3 - 13]$ fb  at $\sqrt{s}= 13$ TeV. 
Using this and Eq (\ref{eq:gsigma}), the inferred value of the dimensionful coupling $g_{s \gamma}$ lies in the range ($5.2 \times 10^{-3},1.1 \times 10^{-2}$)

 At the microscopic level, the interaction of the scalar with photons is expected to be induced by loop involving other particles. In this work, we assume those particles to be fermions.
The effective dimensionful coupling $g_{s \gamma}$ can then be evaluated from first principles and for a loop involving Dirac fermions 
in fundamental representation of $SU(3)$, coupled to the scalar $S$ with Yukawa coupling $y_M$ and having electric charges $Q_f$ and masses $m_f$,
is given by \cite{Gunion:1989we}
\begin{equation} \label{eq:tmp}
g_{s \gamma} = \sum_{f} \frac{\alpha}{4 \pi}  ~ \left( \frac{2 N_c ~y_M ~Q^{2}_f }{m_f} \right) ~ A_{1/2} (T_f) \; ,
\end{equation} 
where, the summation is over all such fermions, $\alpha$ is the fine structure constant, $N_c=3$, $T_f = 4m_f^2/m_s^2$, $A_{1/2}(x) = 2x [ 1 + (1-x) f(x) ]$ and $f(x)$ is of the form

\[
f(x)=
\begin{cases}
 \left( \sin^{-1}\left( \frac{1}{\sqrt{x}} \right) \right)^2 , & x \ge 1, \\
 -\frac{1}{4} \left( \ln \left( \frac{1+\sqrt{1-x}}{1-\sqrt{1-x}} \right) - i \pi \right)^2 , & x < 1 \; .
\end{cases}
\]

We are considering the scalar to be SM singlet and fermions to be vector-like. The CMS collaboration has analysed the data collected at LHC at $\sqrt{s}=8$ TeV for the search of a pair production of charge $2/3$ vector-like  fermions and reported  the lower bound on the mass of the same  to be around $0.85$ TeV \cite{Khachatryan:2015oba}. On the other hand, the analysis performed by the same collaboration using the data collected at  $\sqrt{s}=13$ TeV for the search of pair-produced $5/3$ charged top partners  gave a lower bound on the mass of the same to be $0.95$ TeV  \cite{CMS:2015alb}.  Also, recently the ATLAS collaboration has analysed the data collected at LHC at  $\sqrt{s}=8$ TeV for the search of  singly produced charge $2/3$~{\rm or} (-$4/3$) vector-like fermions and provided the lower bound on the mass of the same to be around $0.95$ TeV \cite{Aad:2016qpo}. In view of this,  we consider the mass of vector-like fermion to be 1.2 TeV in our analysis. It is easy to see that, in eq (\ref{eq:tmp}) if one assumes  $m_f = 1.2$ TeV, one gets 
$T_f = 10.24$ and $A_{1/2} (T_f) = 1.365$. Thus, we find that
\begin{equation} \label{eq:gs-ym}
 \frac{g_{s \gamma}}{({\rm TeV})^{-1}} = ( 3.963 \times 10^{-3} )~ ~ y_M ~Q^2 \;,
\end{equation}
where $Q^2 =\sum Q^{2}_f$. 
Comparison with Eq (\ref{eq:gsigma}) tells us that the range of values of $y_M Q^2$ which is compatible with observations
lies in the range (1.32, 2.73).

\begin{figure}[tbp]
\centering 
\includegraphics[width=.495\textwidth]{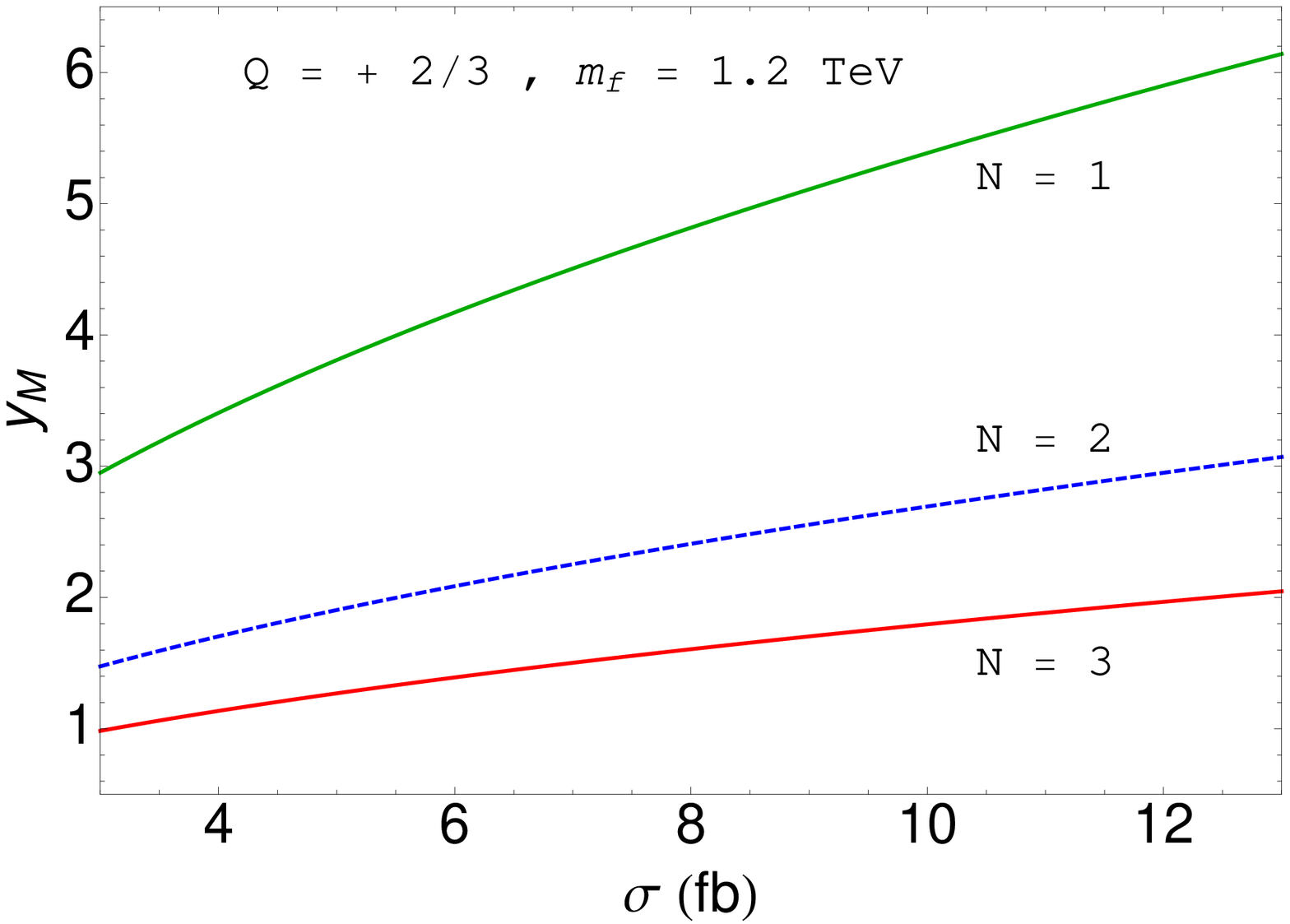}
\hfill
\includegraphics[width=.495\textwidth]{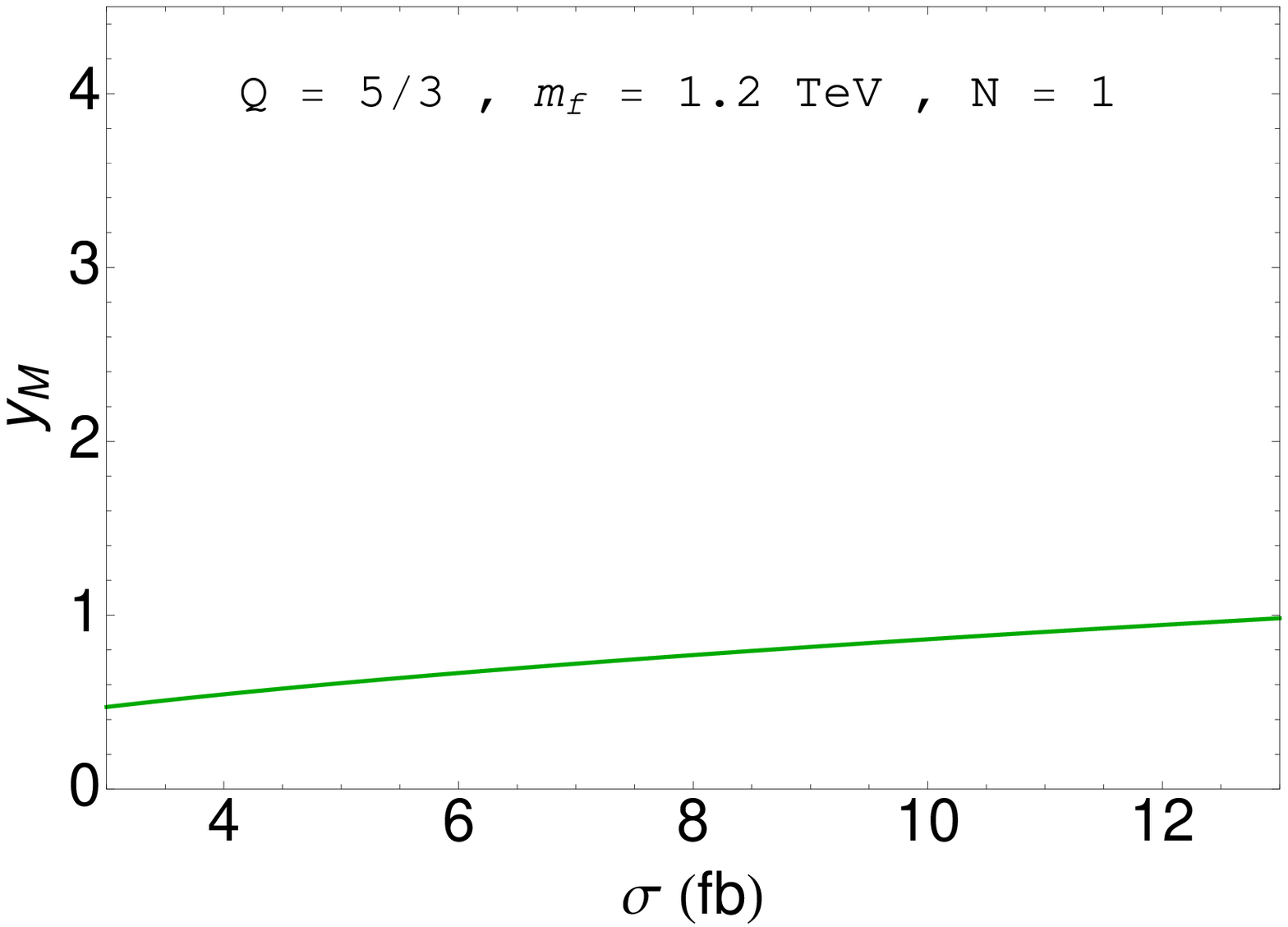}
\caption{\label{fig:i} The calculated values of Yukawa coupling $y_M$ as a function of the cross-section 
$\sigma(pp \rightarrow \gamma \gamma)$ for vector-like fermions with up-type quark charges (top) and for vector-like 
fermions with exotic charges (bottom). Notice the difference in range of $y_M$ values in the two plots. It is easy to see 
that exotic charges easily restore perturbativity while increasing the number of fermions has a more mild effect on 
perturbativity.}
\end{figure}

In  fig (\ref{fig:i}), we have shown the range of values of Yukawa coupling $y_M$ which give cross-section 
$\sigma(pp  \rightarrow \gamma\gamma)  \sim [3-13]$ fb by considering vector-like fermions (with up-type quark charges) in the loop
 with $m_f = $ 1.2 TeV. It turns out that by considering single vector fermion of electric charge $Q =2/3$, 
the theory is strongly coupled for $\sigma(pp  \rightarrow \gamma\gamma)  \ge 4.5$ fb at the scale $m_f$ itself. 
If we have N identical fermions with $Q = 2/3$ in fig (\ref{fig:i}), one might think that one could restore the perturbativity for 
$\sigma(pp  \rightarrow \gamma\gamma)  \sim [3-13]$ fb by considering $N = 3$. 
As can be checked explicitly, though this renders the Yukawa coupling sufficiently small at TeV scale, the theory still 
becomes non-perturbative as we approach higher energies. We will come back to this in sec IV. 
Thus it is a non-trivial task to explain the diphoton excess without running into non-perturbative regime at the TeV scale 
or some sub-Planckian scale.

An alternative option is to consider fermions running in the loop to have larger values of the electric charge \cite{Pilaftsis:2015ycr}. 
It is easy to see from the plot at the bottom in fig (\ref{fig:i}) that for $Q = 5/3$, the corresponding Yukawa coupling remains small 
for the entire range of $\sigma(pp  \rightarrow \gamma\gamma)  \sim [3-13]$ fb and $N = 1$ at least at the TeV scale. 

Before we proceed, let us list the various constraints which we would like to satisfy:
\begin{enumerate}
 \item The cross section $\sigma (pp \rightarrow S \rightarrow \gamma \gamma)$ must lie in the observed range of $3 - 13$ fb,
 \item Perturbativity of all the couplings e.g. Yukawa, scalar couplings, gauge couplings:
All couplings must be sufficiently small so that the theory stays perturbative till Planck scale, 
 \item Vacuum stability:  
 At all scales the scalar couplings must be such that the scalar potential $V$ should always be above its value for the Standard Model vacuum,
  \item mixing angle between Higgs and the singlet scalar must be such that no observable effect takes place,
 \item Given that there is going to be another scalar in the theory apart from SM Higgs, we would like to check whether one can get 
  successful inflation in such scenarios.
\end{enumerate}

In the next section, we discuss how by considering vector-like fermions with an exotic charge $Q > 2/3$, the theory can be kept  
perturbative even upto Planck scale for a reasonable choice of parameters.

\section{A specific scenario}

The gauge group of our model is assumed to be that of the Standard Model (SM) i.e. $SU(3)_c \times SU(2)_L \times U(1)_Y$. 
The bosonic fields in our model are: the SM gauge fields, the SM Higgs field and a real parity even scalar field $\chi$ (assumed to be a singlet).
We consider two models which we shall call model X and model Y and which differ in their fermionic content. 
In the fermionic sector of model X, there is, in addition to the SM fermions, a single colored vector-like fermion, $\Psi$, 
with exotic electric charge which is a singlet under $SU(2)_L$.
On the other hand, in model Y, apart from the SM fermions, there is colored vector-like fermion which is a doublet under $SU(2)_L$. The reason for choosing $SU(2)_L$ doublet fermion is that it allows mixing of exotic charged vector-like fermions with SM quarks, therefore causing decay of the same into SM particles (see sec IV.3 for details).
 
\subsection{Constraints from the scalar sector}

The bosonic sectors of the two models are the same. In particular, in the scalar sector, the real scalar field $\chi$ does not couple to any 
of the gauge fields while the Higgs field couples to the $SU(2)_L$ and $U(1)_Y$ gauge fields. The scalar potential is given by

\begin{eqnarray}
 V(\Phi,\chi) &=& \lambda_H \left( \Phi^\dagger \Phi - \frac{v^2}{2} \right)^2 + \lambda_S \left( \chi^2 - u^2 \right)^2 \nonumber \\
 && + \lambda_{HS} \left( \Phi^\dagger \Phi - \frac{v^2}{2} \right) \left( \chi^2 - u^2 \right) \; ,
\end{eqnarray}
so that the potential is specified by five parameters: $u,v,\lambda_H, \lambda_S, \lambda_{HS}$. 
Expanding the fields about the vev 
\begin{equation}
 \Phi = \frac{1}{\sqrt{2}} \left( 0 ~~~ v + h \right)^T, ~~ \chi = u + s \; ,
\end{equation}
the terms quadratic in the fields $h$ and $s$ are 
\begin{equation}
V_2 = ( \lambda_H v^2 ) h^2 + ( \lambda_S u^2 ) s^2 + ( \lambda_{HS} u v ) s h \; ,
\end{equation}
so that the mass eigenstates $H$ and $S$ will be different from the fields $h$ and $s$ and there will be a mixing between the two. 
The mixing angle $\theta$ and the mass eigenvalues $m_H$ and $m_S$ can be readily determined 
from the parameters specifying the potential (see e.g. \cite{Xiao:2014kba} for details).
The mixing between Higgs and new singlet S is given by \cite{Xiao:2014kba}
\begin{equation}
\theta = \frac{1}{2} \tan^{-1}\bigg( \frac{\lambda_{SH} u~ v}{\lambda_S u^2 - \lambda_H v^2} \bigg) \; .
\end{equation}
To avoid the tight constraints from the LHC on the Higgs decay modes, it a good idea to suppress the mixing between $S$ and $H$. 
We thus need to ensure that the parameters $\lambda_H, \lambda_S, \lambda_{HS}$ are such that for $u \sim {\cal O} $ (TeV), 
$\theta$ turns out to be less than 0.15 \cite{Falkowski:2015iwa} and $m_H$, $m_S$ take up the expected values.

As our scenario involves a heavy scalar, at the scale $m_S$, we can integrate out this heavy scalar and hence solve the equations of motion 
of the theory to obtain a description of the potential in terms of the field $\Phi$ alone. This causes a discontinuity in $\lambda_H$ at the scale 
$m_S$ so that the  new value of the quartic coupling is identified with $\lambda_{H}^{SM}$, the Higgs self-coupling in SM
\cite{EliasMiro:2012ay} i.e. at the scale $m_S$,
\begin{equation} \label{eq:jump}
 \lambda_H - \frac{\lambda_{SH}^2}{4 \lambda_S} = \lambda_{H}^{SM} \; .
\end{equation}

\subsection{Constraints from fermionic sector}

In both models, the newly added fermions are colored so that they couple to the gluons. 
In addition to this, in model X, the fermions couple to the $U(1)_Y$ gauge field while in model Y, the fermions couple to the gauge fields of both 
$SU(2)_L$ and $U(1)_Y$. This ensures that in model Y, we'll have terms such as $g_2 {\bar \psi}_2 \gamma^\mu W_\mu^- \psi_1$ and 
$g_2 {\bar \psi}_1 \gamma^\mu W_\mu^+ \psi_2$. The presence of these terms plays a crucial role in this model.

Now let us turn to the Yukawa terms in the two models. In what follows, the quarks always refer to the third generation quarks of the SM.
In model X, the Yukawa term of the form $\chi {\bar \psi}_L \psi_R$ is possible irrespective of the charge of the fermion.  Moreover, the terms $\chi {\bar \psi}_L u_R$ and $\Phi^c {\bar Q}_L \psi_R$ are also possible if the charge of the fermion is  chosen to be 2/3.  Similarly, the terms $\chi {\bar \psi}_L d_R$ and $\Phi {\bar Q}_L \psi_R$ are also possible if the charge of the fermion is  chosen to be -1/3. 

In model Y, we could denote the doublet fermion $\psi$ by $\psi = (\psi_1 ~~~ \psi_2)^T$.
Since the hypercharge of all the fields which are part of the same multiplet is the same, it is easy to see that $Q_1 - Q_2 = 1$.
In model Y, the Yukawa term of the form $\chi {\bar \psi}_L \psi_R$ is possible irrespective of the hypercharge of the fermion.  Moreover, the terms $\Phi {\bar \psi}_L d_R$, $\Phi^c {\bar \psi}_L u_R$ and $\chi {\bar Q}_L \psi_R$ are also possible if the  hypercharge of the doublet fermion is chosen to be 1/6. In this case the electric charges of the particles in the doublet are  $Q_1 = 2/3$ and $Q_2 = -1/3$, the same as that in the SM quark doublet.  Similarly, the term $\Phi {\bar \psi}_L u_R$ is possible if the hypercharge of the doublet is 7/6, this gives  $Q_1 = 5/3$ and $Q_2 = 2/3$. Finally, the term $\Phi^c {\bar \psi}_L d_R$ is possible if the hypercharge of the doublet is -5/6, which leads to  $Q_1 = -1/3$ and $Q_2 = -4/3$.

The list of all possible Yukawa couplings with different hypercharges in both the models is summarized in Table I.
\begin{table}
\begin{center}
\label{ref:table}
\begin{tabular}{|c|c|c|c|} 
 \hline
Model & Yukawa &  Hypercharge & Vector-like  \\ 
  & terms &  $Y_f$  & fermion content  \\ 
\hline \hline
& & & \\
 &  $\chi {\bar \psi}_L \psi_R$ &  &  \\ 
 & $\chi {\bar \psi}_L u_R$ & $2/3$ & $\psi_{2/3}$ \\
 & $\Phi^c {\bar Q}_L \psi_R$ & & \\
  & & & \\ 
X & $\chi {\bar \psi}_L \psi_R$  & &  \\ 
&  $\chi {\bar \psi}_L d_R$ &  -$1/3$ & $\psi_{-1/3}$ \\
&  $\Phi {\bar Q}_L \psi_R$ & &\\
& & & \\
   & $\chi {\bar \psi}_L \psi_R$  & Exotic/Arbitrary & $\psi_{(Q > 2/3)}$ \\ 
   & & ($Y_f > 2/3$) & \\
   & & & \\
  \hline
   & & & \\
  & $\chi {\bar \psi}_L \psi_R$ & &  \\ 
  & $\Phi {\bar \psi}_L d_R$ &  1/6  & $\left(\psi_{2/3} ~~~ \psi_{-1/3} \right)^T$ \\
  & $\Phi^c {\bar \psi}_L u_R$ & &\\
  &  $\chi {\bar Q}_L \psi_R$ & &\\
     & & & \\
Y & $\chi {\bar \psi}_L \psi_R$   &7/6 & $\left(\psi_{5/3} ~~~ \psi_{2/3} \right)^T$ \\
& $\Phi {\bar \psi}_L u_R$ & & \\
   & & & \\
  &  $\chi {\bar \psi}_L \psi_R$  &-5/6 & $\left(\psi_{-1/3} ~~~ \psi_{-4/3} \right)^T$ \\
  &  $\Phi^c {\bar \psi}_L d_R$ & & \\
 \hline
\end{tabular}
 \caption{Possible Yukawa terms in the Lagrangians of model X and model Y and the corresponding hypercharge of the newly added vector-like fermion.}
\end{center}
\end {table}
It is easy to verify that this exhausts all the possibilities for BSM Yukawa terms in both model X and model Y. 
The arguments of section II suggest that the charge of the newly added fermion must be exotic. 
This means that the Yukawa part of the Lagrangian in model X is 
\begin{equation}
{\cal L}^{\rm X}_{\rm Yuk} = {\cal L}^{\rm SM}_{\rm Yuk} + y_M \chi {\bar \psi}_L \psi_R + {\rm h.c.} \;,
\end{equation}
while the Yukawa part of the Lagrangian for the case with 7/6 hypercharge in model  Y is
\begin{equation}
{\cal L}^{\rm Y}_{\rm Yuk} = {\cal L}^{\rm SM}_{\rm Yuk} + y_M \chi {\bar \psi}_L \psi_R + y_f \Phi {\bar \psi}_L u_R + {\rm h.c.} \; ,
\end{equation}
and the Yukawa part of the Lagrangian for the case with -5/6 hypercharge in model  Y is
\begin{equation}
{\cal L}^{\rm Y}_{\rm Yuk} = {\cal L}^{\rm SM}_{\rm Yuk} + y_M \chi {\bar \psi}_L \psi_R + y'_f \Phi^c {\bar \psi}_L d_R + {\rm h.c.} \;.
\end{equation}


\subsection{Issues of perturbativity: gauge couplings}

By now it is clear that we shall be interested in scenarios in which vector-like fermions with exotic charges are added to the SM.
Each gauge coupling $g_i$ evolves as (see appendix)
\begin{equation}
 \frac{d g_i}{d t} =  \frac{b_i}{(4 \pi)^2} g_i^3 \; ,
\end{equation} 
where $b_i$ are numerical factors.
Just like in the SM, in both model X and model Y,  both $b_3$ and $b_2$ are negative, so, $g_3$ and $g_2$ become small at large energies.
In the models we are considering

  \begin{equation}\label{eq:b1-explicit}
    b_1=
    \begin{cases}
      \frac{41}{10} + \frac{12}{5}Y^2, & \text{model X} \; , \\
      \\
      \frac{41}{10} + \frac{24}{5}Y^2, & \text{model Y} \; ,
    \end{cases}
  \end{equation}
is a positive quantity. The value of $b_1$ is larger in both the models $X$ and $Y$ as compared to the SM in which $b_1 =  4.1$.
If the coupling $g_1$ at the scale $\mu_i$ is $g_1(\mu_i)$, then $g_1$ at a higher scale $\mu$ is 
\begin{equation} \label{eq:g1-sol}
 g_1 (\mu) = \frac{g_1(\mu_i)}{\left( 1 - \frac{b_1 {g_1(\mu_i)}^2 }{8 \pi^2} \ln \left( \frac{\mu}{\mu_i} \right) \right)^{1/2}} \; ,
\end{equation}
so that, at the scale 
\begin{equation}
  \mu = \mu_i \exp \Bigg( {\frac{8 \pi^2}{b_1 {g_1(\mu_i)}^2}} \Bigg)\; ,
\end{equation}
the coupling $g_1$ hits a Landau pole. For SM, this happens at $\mu \approx 10^{41}$ GeV (in one-loop approximation).
In the scenarios we are interested in, larger value of $b_1$ turn up and the Landau pole is found at a lower scale. 

Now, one may wish to have inflation with the scalars present in the model. 
The potential at large values of the fields is quartic and this leads to large field inflation.
During large field inflation, the inflaton field undergoes super-Planckian excursion so that to get the minimum $60$ e-foldings of 
inflation, the field excursion $\Delta \phi \sim {\cal O} (10)  M_{\rm Pl}$, where, $M_{\rm Pl}$ is the reduced Planck mass. 
Thus, if we wish our perturbative calculations to be valid and hence trustworthy during inflation, we'd want all the couplings to stay 
small till super-Planckian scales. This could also help in keeping the inflaton potential sufficiently flat.
Thus, we impose the constraint that the maximum value a growing coupling should take must be 
$\sqrt{4 \pi}$ at $\mu = 1.2 \times10^{20}$ GeV $\approx 50 M_{\rm Pl}$ so that perturbativity (in matter sector) 
breaks down at scales $\mu \ge 50 M_{\rm Pl}$ only.
Of course, at $\mu \sim M_{\rm Pl}$, gravity is expected to become strongly coupled but this is a separate subject in itself.

Let us now impose this constraint: we would like to have
\begin{equation} \label{g1_constraint}
 g_1 (\mu) < \sqrt{4 \pi} ~{\rm for}~ \mu < 1.2 \times 10^{20} {\rm GeV}.
\end{equation}
Assuming the mass of the fermion to be 1.2 TeV, using the fact that $g_1$ at the scale $1.2$ TeV is 0.47 and 
using Eq (\ref{eq:b1-explicit}), we find that the corresponding allowed maximum value of $Y^2$ is
\begin{equation} \label{eq:qsqn}
  Y_{\rm max}^2 =
    \begin{cases}
      2.08, & \text{model X} \; , \\
      1.04, & \text{model Y} \; .
    \end{cases}
\end{equation}
This tells us that if we impose this constraint (and assume $m_f = 1.2$ TeV), then, in model Y, the case with 
Y=7/6 (i.e. $Q_1 = 5/3$ and $Q_2 = 2/3$) is not consistent with our requirements.\footnote{Of course, one could change the mass of the fermion and could argue that hitting Landau pole around Planck scale is 
not undesirable. In that case one could work with Y=7/6 fermion as is done in \cite{Hamada:2015skp}. }
 Therefore the case which will interest us most will be the one in which the hypercharge of the fermion doublet is chosen to be  $-5/6$.
 
\begin{figure}
\begin{center}
\includegraphics[width=0.45\textwidth]{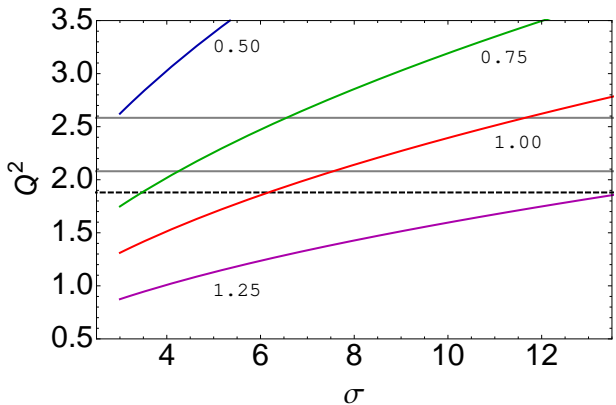}
\end{center}
\caption{The uppermost horizontal line here is the value of $Q^2$ one gets by imposing the requirement that $g_1$ is less than 
$\sqrt{4 \pi}$ for scales below $10^{20}$ GeV (see text for details) for the mass of the fermion fixed to be 1.2 TeV in model Y.
The middle horizontal line is the value of $Q^2$ one gets by imposing the same requirement  in model X. 
The lowest horizontal line is the value of $Q^2$ for the hypercharge $Y =-5/6$ for the multiplet in model Y. 
The curves correspond to the $Q^2$ found from Eq(\ref{eq:gs-ym}) for the values of the Yukawa coupling $y_m$ shown.
For a given $\sigma (pp \rightarrow S \rightarrow \gamma \gamma)$, this means that there 
is a unique allowed value of the Yukawa coupling $y_m$.
}
\label{fig:QsqN-vs-sigma}
\end{figure}

Continuing with the assumption that the mass of the fermion is 1.2 TeV, eq (\ref{eq:qsqn}) must be compared with what one obtains by combining 
eq (\ref{eq:gsigma}) and eq (\ref{eq:gs-ym})
\begin{equation}
 Q^2  = \frac{1}{1.32~ y_m}  \left( \frac{\sigma}{\rm fb} \right)^{1/2} \; ,
\end{equation}
where, $Q^2= Q_1^2 + Q_2^2$ so that 
\begin{equation}
 Q^2 = 2 Y^2 + \frac{1}{2} \;.
\end{equation}
In fig (\ref{fig:QsqN-vs-sigma}), we plot $Q^2$ against the 
allowed range of cross section $\sigma (pp \rightarrow \gamma \gamma)$, for various choices of the Yukawa coupling $y_m$.
Imposing the additional requirement of Eq (\ref{eq:qsqn}) ensures that for a given value of cross-section, there is a unique
value of the Yukawa coupling (for $m_f$ fixed at 1.2 TeV). 

The most important lesson to be learnt from fig (\ref{fig:QsqN-vs-sigma}) is that for Model X, the Yukawa coupling $y_M$
consistent with (i) $\sigma > 3$ fb, (ii) not having Landau pole in $g_1$ below $10^{20}$ GeV (assuming $m_f = 1.2$ TeV) is 
such that $y_M > 0.63$. Similarly, for model Y, the Yukawa coupling consistent with these requirements is $y_M > 0.5$.
Since the Yukawa coupling can not be kept arbitrarily small, maintaining perturbativity till Planck scale is a fairly non-trivial constraint.

In summary, if eqn (\ref{eq:gs-ym}) was the only constraint, then we could have decreased 
$y_M$ as much as we liked by increasing the quantity $Q^2$ but this quantity is fixed by Eq (\ref{eq:qsqn}) and so we no longer 
have that freedom.
Secondly, if some of the values of the Yukawa coupling in this allowed range are somehow not viable, 
then there is no scope for explaining the corresponding value of cross-section from this model while maintaining perturbativity
and $m_f = 1.2$ TeV.

\begin{figure}[tbp]
\centering 
\includegraphics[width=.45\textwidth]{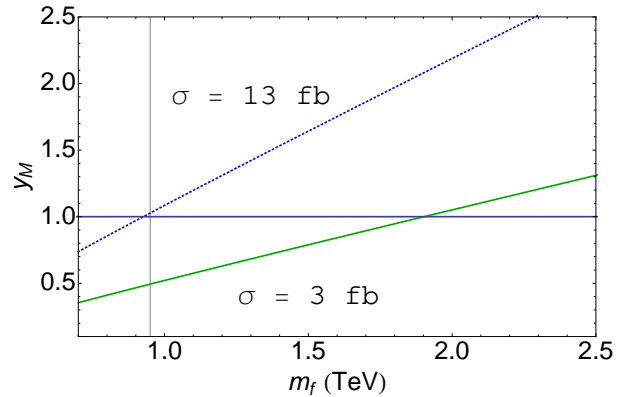}
\caption{\label{fig:changing_fermion_mass}
 The inferred value of Yukawa coupling for Model Y at the scale $m_f$ for any chosen $m_f$ (the mass of the fermion) 
and for the two extreme values of the diphoton rate obtained
with hypercharge of the fermion doublet chosen to be $-5/6$ i.e. within the limits specified by Eq (\ref{eq:qsqn}).
We find that having lighter fermion helps in keeping the coupling small. Though for values of $m_f$ below 950 GeV (i.e. left of the vertical line), all cross sections of the diphoton signal can be explained, these values are below the lower limit on the mass of vector like fermion of charge -$4/3$ obtained by the LHC (see \cite{Aad:2016qpo}).}
\end{figure}

 Let us now see what happens when we change the mass of the fermions in the doublet 
$(\psi_{-1/3}~~~ \psi_{-4/3})^T$ considered in model Y. 
As we saw in section II (in the discussion just before eq (\ref{eq:gs-ym})), 
the experimental lower limit on the mass of vector like fermions of charge $-4/3$ has been obtained by the ATLAS collaboration and is around 0.95 TeV \cite{Aad:2016qpo}. Using eq (\ref{eq:gsigma}) and eq (\ref{eq:tmp}) along with the requirement imposed by eq (\ref{g1_constraint})
ensures that in this scenario, increasing the mass of fermion increases the inferred value of Yukawa coupling of the fermion 
in order to match the observed diphoton rate. 
This is illustrated in fig (\ref{fig:changing_fermion_mass}) which also shows that if $m_f > 1.7$  TeV, no allowed value of diphoton 
rate can be explained while keeping the Yukawa coupling smaller than one. 
From the RG equation of $\lambda_S$ (see appendix), it is worth noticing that having a value of $y_M > 1$ can be an impediment in maintaining perturbativity till Planck scale since in such a case $\lambda_S$ and $\lambda_{SH}$ can undergo 
huge RG evolution. 
 From fig (\ref{fig:changing_fermion_mass}), we also learn that if $m_f < 0.9$ TeV, all the allowed 
values of the diphoton rate can be explained with Yukawa couplings smaller than unity. 
However, this small value of $m_f$ is not favoured by the ATLAS results \cite{Aad:2016qpo}.

\section{Perturbativity, vacuum stability and other constraints}

In order to analyse the issues of perturbativity and vacuum stability, we need to find the Renormalization Group (RG) evolution equations for
the scenario we are considering. 
The one-loop RG evolution equations for a model with SM particles, a real singlet scalar and  $SU(2)_L$ singlet vector-like quark (i.e. $Q = 2/3$)
have already been found in Ref. \cite{Xiao:2014kba}. 
In order to deal with model X and model Y, we need to generalise the equations presented in \cite{Xiao:2014kba} to take into account  both $SU(2)_L$ singlet and $SU(2)$ doublet vector-like fermions with exotic charges i.e. $Q > 2/3$. 
 
\subsubsection{Solution of RGEs: Perturbativity} \label{sec:soln}
The detailed RG evolution equations for both model X and model Y have been given in appendix \ref{App:AppendixA}.  
We numerically solve the RG equations for three different regimes: 
(i) from $m_t$ (= 173.1 GeV) to $m_S$ (= 750 GeV),
(ii) from $m_S$ to $m_f$ (the mass of the fermion), and, 
(iii) from $m_f$ to Planck scale and beyond. 
In regime (i), one works with the RG equations of the SM ($g_1$, $g_2$, $g_3$, $y_t$ and $\lambda_H^{SM}$) while in 
regime (ii), two new scalar couplings ($\lambda_S$ and $\lambda_{SH}$) enter the description.
The threshold corrections due to the existence of the singlet scalar $S$ with mass $750$ GeV causes the  
familiar discontinuity in the evolution of $\lambda_H$ \cite{EliasMiro:2012ay} as given in eq (\ref{eq:jump}).
Finally, in regime (iii), in model X, we have Yukawa coupling $y_M$ and in model Y, we have Yukawa couplings $y_M$ and $y_F$.

\begin{figure}
\begin{center}
\includegraphics[width=0.45\textwidth]{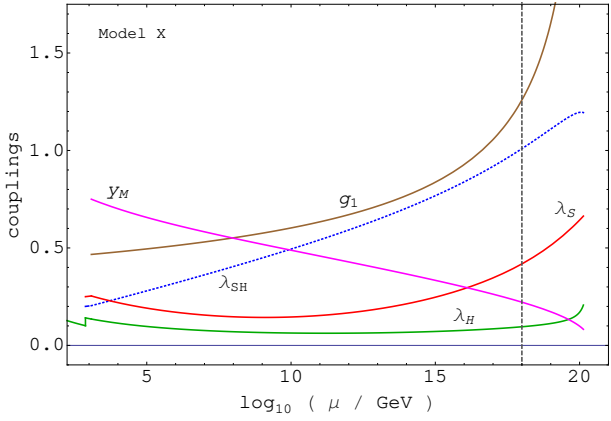}
\end{center}
\caption{For model X: the evolution of the various couplings for the case $y_M (m_f) = 0.75$, $\lambda_S (m_s) = 0.25$ and $\lambda_{SH} = 0.20$. 
Here $m_f = 1.2$ TeV, and this choice of parameters can explain the diphoton excess if 
$\sigma (pp \rightarrow S \rightarrow \gamma \gamma) \approx 4$ fb.
The square of the electric charge (and hence hypercharge) of the fermion is chosen to be $Q^2 = Q_{\rm max}^2 =  2.08$ (see Eq (\ref{eq:qsqn})). 
Notice that $g_1$ is below $\sqrt{4 \pi}$ for $\mu < 1.2 \times 10^{20}$ GeV.
The region to the right of the vertical line corresponds to $\mu > 10^{18}$ GeV where quantum gravitational effects are expected to be important.}
\label{fig:lambdas}
\end{figure}

\begin{figure}
\begin{center}
\includegraphics[width=0.45\textwidth]{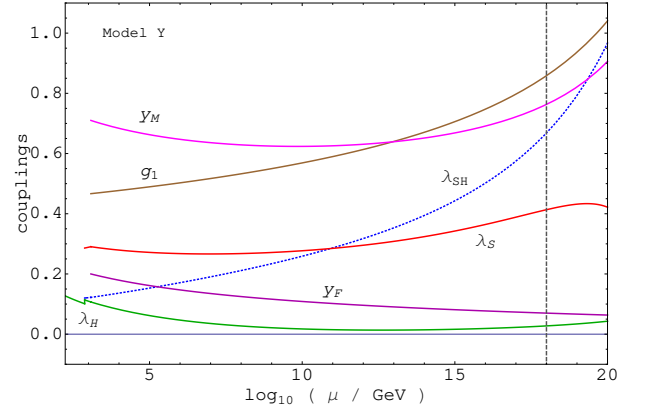}
\end{center}
\caption{For model Y: the evolution of the various couplings for the case $y_M (m_f) = 0.71$, $\lambda_S (m_s) = 0.285$ and $\lambda_{SH} = 0.12$. 
Here $m_f = 1.2$ TeV, and this choice of parameters can explain the diphoton excess if 
$\sigma (pp \rightarrow S \rightarrow \gamma \gamma) \approx 3$ fb.
 The hypercharge of the fermion doublet is chosen to be $Q =  -5/6$ (see Eq (\ref{eq:qsqn})). 
Notice that $g_1$ is below $\sqrt{4 \pi}$ for $\mu < 1.2 \times 10^{20}$ GeV.
The region to the right of the vertical line corresponds to $\mu > 10^{18}$ GeV where quantum gravitational effects are expected to be important.}
\label{fig:lambdas_Y}
\end{figure}

In fig (\ref{fig:lambdas}), we have shown the RG evolution of the scalar couplings $\lambda_H$, $\lambda_S$ and $\lambda_{SH}$, 
the gauge coupling $g_1$ and the Yukawa coupling $y_M$ for model X for a choice of parameters mentioned in the figure caption.
This particular choice of parameters can easily explain the cross-section $\sigma \approx 4$ fb using model X. 
Similarly, in fig (\ref{fig:lambdas_Y}), we have shown the RG evolution of the scalar couplings $\lambda_H$, $\lambda_S$ and $\lambda_{SH}$, 
the gauge coupling $g_1$ and the Yukawa couplings $y_M$ and $y_F$ for model Y for the choice of parameters mentioned in the figure caption.
This particular choice of parameters can easily explain the cross-section $\sigma \approx 3$ fb using model Y. 
 It is noteworthy that, as can be guessed from fig (\ref{fig:changing_fermion_mass}), 
all observed values of the diphoton rate can be explained by considering a light enough fermion while maintaining perturbativity
till $\mu \sim 50 M_{\rm Pl}$ by appropriately choosing the $\lambda_S$ and $\lambda_{SH}$ at the scale of mass of the 
singlet scalar. Also, as is clear from fig (\ref{fig:lambdas}) and fig (\ref{fig:lambdas_Y}), we evolve the RG equations till well beyond Planck scale.
It is noteworthy that the couplings $\lambda_H$, $\lambda_S$ and $\lambda_{SH}$ are such that not only is the mixing of $S$ with SM Higgs 
$H$ negligible, but also, the mass eigenvalues take up the correct values $m_S$ and $m_H$.

On the other hand, what we have shown is that both model X and model Y stay weakly coupled (since all the couplings stay smaller than unity) 
all the way till Planck scale while still explaining the observed diphoton excess.

\begin{figure}
\begin{center}
\includegraphics[width=0.45\textwidth]{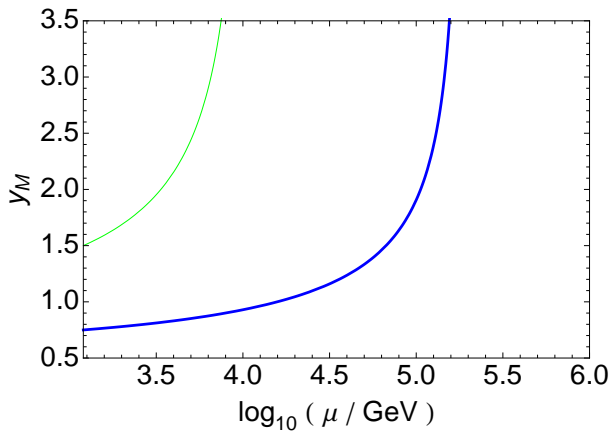}
\end{center}
\caption{RG evolution of Yukawa coupling $y_M$ for two cases with multiple copies (N) of vector-like fermion in model X: (i)  The solid (green) line shows the evolution of  $y_M$ for the parameters $y_M(m_f)=1.5$, $N=2$, $Q=2/3$ and $\sigma (pp \rightarrow S \rightarrow \gamma \gamma) \approx 3$ fb, (ii)  The thick (blue) line shows the evolution of $y_M$  for the parameters $y_M(m_f)=0.75$, $N=4$, $Q=2/3$ and $\sigma (pp \rightarrow S \rightarrow \gamma \gamma) \approx 3$ fb.}
 \label{fig:multipleN}
\end{figure}

  \paragraph*{\bf { \emph {Perturbativity issue with multiple generations of up-type vector-like fermion:}}}
Let us discuss the situation if instead of considering vector-like fermions with exotic charges, we consider vector like fermions with  up-type quark  charge but increase the number of flavours/generations of the fermions. If we consider $N_f$ vector-like fermions,  the eq (\ref{eq:b1-explicit}) will be modified to
\begin{equation}\label{eq:b1-explicit_multN}
    b_1=
    \begin{cases}
      \frac{41}{10} + \frac{12}{5} N_f Y^{2}_f, & \text{model X} \; , \\
      \\
      \frac{41}{10} + \frac{24}{5} N_f Y^{2}_f, & \text{model Y} \; ,
    \end{cases}
  \end{equation}
where $Y_f$ represents the hypercharge of vector-like fermion. Similarly,  the equation (\ref{eq:qsqn}) representing the maximum allowed value of hypercharge from the Landau pole requirement will be modified to. 
\begin{equation} \label{eq:qsqn_multN}
  {(N_f Y^{2}_f)}^{\rm max}  \sim
    \begin{cases}
 2.08 & \text{model X} \; , \\
  1.04 & \text{model Y} \; .
    \end{cases}
\end{equation}
Incorporating $Y_f=2/3$ for a $SU(2)$ singlet up-type vector-like fermion in model X and $Y_f=1/6$ for a $SU(2)$ doublet up-type vector-like fermion in model Y, we obtain
\begin{equation} \label{eq:N_multN}
  N^{\rm max}_f \sim
    \begin{cases}
 4 & \text{model X} \; , \\
  37 & \text{model Y} \; .
    \end{cases}
\end{equation}
We analyse two cases of model X in which one can explain at least the minimum value of diphoton cross-section ($\sigma (pp \rightarrow S \rightarrow \gamma \gamma) \approx 3$ fb) by considering multiple up-type vector-like fermions and  particular value of Yukawa couplings. 
In those cases, we consider (i) $y_M=1.5$, $N_f=2$ and (ii)  $y_M=0.75$, $N_f=4$. 
As is clear from the fig (\ref{fig:multipleN}), the Yukawa coupling become large at a very low energy scale in both the cases. 
The reason behind this is the competition between the positive contribution coming from the term of the type $N_f y^{2}_M$ and the negative contribution coming from the term of the type $ Q_f g^{2}_1$ in the RG evolution equation of the Yukawa coupling $y_M$ (see equation (\ref{eq:YTyM}) in the appendix). If the former gets dominant (due to large $N_f$), it will make the theory non-perturbative at a low energy scale. 
However,  if the later become dominant by increasing value of $Q_f$, it keeps the theory perturbative till Planck scale. 
Therefore, we show that though increasing number of generation does help in explaining the diphoton cross-section at TeV scale (see sec II), 
it does not allow us to keep the theory perturbative all the way till Planck scale. 
\subsubsection{Vacuum stability and inflation}
In the last section, we found two models compatible with the observed diphoton excess and a regime in which these model stay weakly 
coupled till Planck scale. Let us now find a few more consequences of this.
It is well known that at energy scale above $10^{10}$ GeV, the self coupling of the SM Higgs turns negative 
\cite{Holthausen:2011aa,EliasMiro:2011aa,Buttazzo:2013uya,Goswami:2014hoa}. 
This observation has motivated many scenarios of physics beyond the Standard Model. 
It is well known that adding more fermions typically causes $\lambda_H$ to turn negative at even smaller scales. 
On the other hand, addition of more scalars can potentially enhance $\lambda_H$ so much that it may hit Landau pole below Planck scale. 

Before we discuss electroweak vacuum stability in model X and model Y, it is worthwhile recalling the conditions of stability. At 
large field values, the scalar potential is of the form 
\begin{equation}
 V = \lambda_H (\Phi^\dagger \Phi)^2 + \lambda_S \chi^4  + \lambda_{SH} \chi^2 \left( \Phi^\dagger \Phi \right)  \; .
\end{equation}
From this, it is clear that if all the three couplings $\lambda_H$, $\lambda_S$ and $\lambda_{HS}$ are positive, the potential 
will be positive and hence stable (since the SM vacuum corresponds to $V=0$). 
Had the term $\lambda_{HS}$ been negative, the requirement of obtaining a positive scalar potential for asymptotically large values of 
the fields leads to the conditions: $4 \lambda_{S}\lambda_{H} > \lambda^{2}_{SH}, \  \lambda_{H} > 0, \  \lambda_{S} > 0$.

As can be seen in Fig (\ref{fig:lambdas}) and Fig (\ref{fig:lambdas_Y}), in the present scenario, the self coupling of the SM Higgs ($\lambda_H$) 
stays positive and small till Planck scale. Moreover, the self coupling of the real singlet scalar $\lambda_S$ and the 
coupling between the Higgs and the real singlet scalar ($\lambda_{SH})$ also stay positive and small till Planck scale. 
Thus, in these models, with the choice of parameters we have shown, the vacuum stability problem is solved. 
At this stage, we must mention that the instability scale in SM is $10^{8}$ GeV at one-loop accuracy, $10^{10}$ GeV at two-loop accuracy
and even higher at three-loop accuracy \cite{Buttazzo:2013uya}. In our calculations, we have restricted our analysis 
to one-loop accuracy. Since the instability scale gets pushed up in energy as one increases the accuracy of calculation, 
and since the vacuum gets stabilized by the threshold effects \cite{EliasMiro:2012ay} 
of the newly added scalar (at 750 GeV), one expects the stability to be maintained if one repeats these calculations at two-loops. 
Still, it may be worthwhile to check our conclusions for two-loop accuracy.

The recent combined results from Planck, BICEP and KECK \cite{Ade:2015tva,Array:2015xqh}
experiments tell us that inflation with both $\lambda \phi^4$
potential and $m^2 \phi^2$ potential are ruled out by upper limits on primordial B-mode polarisation of CMB.
Since this is typically the form of the potential at large field values for most particle physics scenarios, this is a very important 
observational constraint. A well known way out is to consider non-minimal coupling of the inflaton to gravity of the form $\xi R \phi^2$
where $R$ is the Ricci scalar. This non-minimal coupling is actually inevitable from the point of view of super-gravity. 
It is well known that such scenarios can easily be made compatible with data.

In SM, the negativity of the self coupling $\lambda_H$ at high scale makes it difficult to use SM Higgs as the inflaton even 
with a non-minimal coupling to gravity (see however \cite{Bezrukov:2014ipa} for a recent attempt). 
In our model, at high scales, the potential in both $H$ direction and $S$ direction is of quartic form with $\lambda \sim {\cal O}(0.1) > 0$. 
It is then obvious that a non-minimal coupling of these fields to gravity of the form $\xi R \phi^2$ with $\xi \sim 10^4$ (from Eq 19 of \cite{Bezrukov:2013fka})
leads to successful inflation. Assuming the number of e-foldings of inflation to be 57.7, one obtains (to lowest order in $1/\xi$),
\begin{equation}
n_s \approx 0.96 ~~\text{and}~~ r \approx 0.003 \; ,
\end{equation}
which are compatible with the observations of Planck.
Notice that since the corresponding non-minimal coupling is large, one may need to worry about quantum corrections such as those 
dealt with in \cite{Salvio:2015kka}. 
Given this, one should re-analyse the RG evolution of all couplings after including the effects of the non-minimal coupling $\xi$ 
(see e.g. \cite{Herranen:2015ima,McDonald:2016cdh}, where this issue is partially addressed). We leave this kind of analysis for future work.

Thus, we have shown that (i) the vacuum stability problem gets solved in both model X and model Y, and, 
(ii) the both the SM Higgs and the 750 GeV real singlet scalar whose decay causes the diphoton excess are very good candidates for 
inflaton in these models. 

\subsubsection{Additional constraints}

Let us now consider an additional constraint which distinguishes between the two models we have been working with.
In model X, in order to explain the diphoton excess and maintain perturbativity, we chose the charge of the fermion to be exotic. 
But there is no decay mode for a particle with this value of charge. This is because, as can be seen from the discussion in section IIIB,
the only way a model X fermion can decay into SM  particles is if its charge is either 2/3 or -1/3. 
If we consider the charge to be arbitrary, it will be stable and may have a dangerous relic abundance.

Given that the fermions are $SU(3)$ charged as well, one may think that the fermions annihilate to negligible abundances because of the strong interactions. 
For the mass of fermion $m_f\sim$ TeV, the relic abundance turns out to around $10^{-4}$ per nucleon. Though this is quite small,
according to \cite{Halyo:1999wq}, the abundance of any particles with charge more than 0.83 must be less than $5 \times 10^{-22}$ per nucleon.
Therefore, it may seem that model X is not viable. 
 
However, since the fermions are colored, the story could be more complicated.  As mentioned in \cite{Kang:2006yd}, for $T<\Lambda_{\rm QCD}$, the colored heavy fermions can form bound states (``pions") made up of these exotic fermions and effectively have an enhanced cross section for annihilation (given by the capture cross section) so that all fermions which form bound states annihilate away. Thus, the relic abundance is given by the fraction of these fermions which did not become bound. Therefore, the overall relic abundance of exotic fermions might be less than the one found above (i.e. $10^{-4}$). 
In this sense the concern about the longevity of the fermion in model X is similar to that of the long-lived gluinos in split-SUSY \cite{ArkaniHamed:2004fb},
we do not address this issue here. 

Instead we argue that the $SU(2)_L$ doublet fermion in model Y decays into SM particles.
 In model Y, since the fermion is $SU(2)_L$ doublet, the Lagrangian contains terms such as
\begin{eqnarray}
&& {\cal L}  = {\cal L}^{\rm SM}_{\rm Yuk} + {\cal L}^{\rm SM}_{\rm gauge} + y_M \chi {\bar \psi}_L \psi_R + y_F \Phi^c {\bar \psi}_L d_R   \nonumber\\
&&  + g_2 {\bar \psi}_2 \gamma^\mu W_\mu^- \psi_1  + g_2 {\bar \psi}_1 \gamma^\mu W_\mu^+ \psi_2. 
\end{eqnarray}
Because of the presence of these terms, $\psi_1$ decays into a bottom quark and Higgs while $\psi_2$ decays to $W^-$, bottom quark and Higgs.
Hence from this point of view, model Y is the preferred model which serves our purpose.

Notice that this requirement of decay of the fermion can be used to rule out not only model X but also all values of the hypercharge in model Y other than $7/6$ and $-5/6$. But since the hypercharge $7/6$ leads to a Landau pole in $g_1$ around Planck scale, we prefer to work with model Y with 
hypercharge of the fermion doublet chosen to be -5/6.

\section{Summary}

One of the most peculiar features of the SM is that it stays perturbative and hence calculable till Planck scale.
This does not hold good for arbitrary variations of the SM.  
In this work, we first began by arguing that if the 750 GeV diphoton excess is interpreted as being due to heavy scalar 
(heavier cousin of the SM Higgs) coupled to the photons through a fermion loop, 
then it is generically difficult to maintain perturbativity as we probe higher energies. 
We did this by demonstrating that the inferred values of the couplings in any scenario which explains the 
diphoton excess are generically large (see fig (\ref{fig:i})). 

We then presented a scenario which requires adding very few new particles and which stays weakly-coupled till Planck scale. 
These new particles are a singlet real scalar (whose decay causes diphoton excess) with mass 750 GeV and 
a coloured vector-like fermion with exotic charge $Q > 2/3$ and mass $m_f$ such that $m_f \sim 1.2 $ TeV.
We studied two variations of this scenario: in model A, the newly added fermion is a singlet under $SU(2)_L$ while 
in model B, the newly added fermion is a doublet under $SU(2)_L$.
We argued that the charge of the newly added coloured fermion can not be that of a quark (i.e. $2/3$ or $1/3$) 
in order to explain the observed diphoton excess if one wishes to maintain perturbativity till high scale.
We then limited the possible hypercharges of the fermions in the two models from the requirement that the gauge coupling $g_1$ stays sufficiently 
small till a little above Planck scale (see fig (\ref{fig:QsqN-vs-sigma})). 
Moreover, we chose the couplings of the scalar sector such that the newly added scalar has negligible mixing with the SM Higgs and the 
mass eigenvalues take up the correct values of the masses of the scalar particles. 
 
 We then showed that using a lighter fermion ($m_f \sim 900$ GeV), 
one can (i) explain the diphoton excess over the entire range 
of $\sigma [3-13]$ fb, and also, (ii) maintain perturbativity till Planck scale. 
Using a heavier fermion ($m_f \sim 1200$ GeV), it is still possible to maintain perturbativity till Planck scale for 
some choice of parameters but it becomes difficult to explain all the possible values of the diphoton production 
cross section (in this context see fig (\ref{fig:changing_fermion_mass})).  
However, as we discussed in section II, the experimental lower limits on the masses of any new vector-like fermions (with various charges) 
are around 0.85-0.95 TeV \cite{CMS:2015alb,Khachatryan:2015oba,Aad:2016qpo}, so, we mostly considered the mass of the fermions to be bigger than
that, around 1.2 TeV. From fig (\ref{fig:changing_fermion_mass}), this value of fermion mass successfully explains only smaller value of the cross-section
i.e. $[3-9]$ fb.

We found the Renormalization Group evolution equations for these models and evolved the couplings with this particle content 
and reasonable choices of parameters.
This helped us in finding a regime in which these models stay weakly coupled till Planck scale, the scalar couplings 
$\lambda_H$, $\lambda_S$ and $\lambda_{SH}$ stay positive i.e. the vacuum stability problem is solved and one obtains successful 
inflation. 
We showed that if instead of adding fermions with a large value of charge, we add multiple copies of a fermion with charge 2/3, we are unable 
to maintain perturbativity all the way till Planck scale.

We then argued that in model X, since the charge of the fermion is large, it does not decay into any SM particles and it can lead to dangerous 
abundance of the newly added fermions in the Universe. Thus, the proposed model Y is the only one satisfying all the constraints we have imposed and is hence the preferred model.
Since the electric charges of the fermion doublet in model Y are -4/3 and -1/3, the recent observational limits of \cite{Aad:2016qpo} help us put a lower limit on the 
mass of the fermions forming the doublet.

\appendix
\section{Renormalization Group Evolution Equations} \label{App:AppendixA}

In this appendix, we present the Renormalization Group Evolution Equations (RGEs) for our scenario.

\subsubsection{Modified RGEs for gauge couplings:}

The RG evolution equation for a coupling is found from 
its beta function, for a gauge coupling $g$, 
\begin{equation}
  \frac{d g}{dt} = \beta (g) \; ,
\end{equation}
where $t = \log (\mu/M_t)$, $\mu$ being the energy scale of interest and $M_t$ being the mass of top quark.

The one-loop beta function for a general $SU(N)$ gauge theory is given by
(see eq (73.41) and eq (78.36) of \cite{Srednicki:2007qs} and eq (6.38) of \cite{Ramond:1999vh}) 
\begin{equation}
 \beta (g) = - \frac{1}{3} \frac{g^3}{(4 \pi)^2} \Big( 11 T(A) - 4 T(R_{DF}) - T (R_{CS})  \Big) + {\cal O}(g^2) \; ,
\end{equation}
where $T(A)$ is the index of the adjoint representation of gauge group, $T(R_{DF})$ is the index of the
representation of any Dirac fermion in the theory while $T(R_{CS})$ is the index of the representation of any
complex scalar in the theory. Recall that if we have Weyl (or Majorana) fermions in theory, they contribute half 
as much as a Dirac fermion. Similarly, if we have real scalars in theory, they contribute half 
as much as a complex scalar. 
Given this, it is clear that the RG equation for $g_3$ in model X is 
\begin{equation}
 \frac{d g_3}{dt} = \frac{g_3^3}{(4 \pi)^2} \bigg( -7 + \frac{2}{3} N_f \bigg) \; ,
\end{equation}
while the RG equation for $g_3$ in model Y is 
\begin{equation}
 \frac{d g_3}{dt} = \frac{g_3^3}{(4 \pi)^2} \bigg( -7 + \frac{4}{3} N_f \bigg) \; ,
\end{equation}
where $-7$ in the braces is the contribution from SM particles alone. There is an extra contribution from $N_f$ vector-
like fermions
but none from the scalar since it is a gauge singlet.
In model X, since none of the extra particles are charged under $SU(2)_L$ of the SM, the RG equation for $g_2$ remains unchanged 
i.e. 
\begin{equation}
 \frac{d g_2}{dt} = \frac{g_2^3}{(4 \pi)^2} \bigg( - \frac{19}{6} \bigg) \; .
\end{equation}
For model Y, the RG equation for $g_2$ is
\begin{equation}
 \frac{d g_2}{dt} = \frac{g_2^3}{(4 \pi)^2} \bigg( - \frac{19}{6} +  \frac{13}{6} N_f \bigg) \; .
\end{equation}

Similarly, the one-loop beta function for a $U(1)$ gauge theory is given by (see eq (66.29) of \cite{Srednicki:2007qs}) 
\begin{equation}
 \beta (g) = \frac{g^3}{(4 \pi)^2} \bigg( \sum_\psi Q_\psi^2 + \frac{1}{4} \sum_\phi Q_\phi^2 \bigg) \; ,
\end{equation}
where $Q_\psi$ is the $U(1)$ charge of a Dirac fermion and $Q_\phi$ is the $U(1)$ charge of a complex scalar.
Using this, we can find the RG equation for $g_1$ in model X
\footnote{Notice that, as is always done, we replace $g_1$ by $\sqrt{\frac{3}{5}} g_1$ in order to be consistent with GUT normalization.}
\begin{equation}
  \frac{d g_1}{dt} = \frac{g_1^3}{(4 \pi)^2} \bigg( \frac{41}{10} + \frac{12 N_f Q_f^2}{5} \bigg) \; .
\end{equation}
The RG equation for $g_1$ in model Y is
\begin{equation}
  \frac{d g_1}{dt} = \frac{g_1^3}{(4 \pi)^2} \bigg( \frac{41}{10} + \frac{24 N_f Q_f^2}{5} \bigg) \; .
\end{equation}
As previously, the first term in the parenthesis viz 41/10 is the contribution of SM field content and 
the second term is due to the extra $N_f$ Dirac fermions of hypercharge $Q_f$.

\subsubsection{Modified RGEs for Yukawa couplings:}

For a Yukawa coupling, the beta function is defined by 
\begin{equation}
 \frac{d y}{d t} = \frac{y}{(4 \pi)^2} \beta (y) \; .
\end{equation}
In SM, the Yukawa couplings take the form of three complex $3 \times 3$ matrices, $Y_u$, $Y_d$, $Y_e$
(where, the matrix $Y_u$ is responsible for giving mass to the up type quarks of all the generations etc).
The one-loop beta functions of Yukawa coupling matrices for SM are given by (see e.q. (6.42) of \cite{Ramond:1999vh})
\begin{eqnarray}
\beta_u^{SM} &=& \frac{3}{2} ( Y_u^\dagger Y_u - Y_d^\dagger Y_d ) + T - \bigg( \frac{17}{20} g_1^2  + \frac{9}{4} g_2^2 + 8 g_3^2 \bigg) \; , \nonumber \\ 
\beta_d^{SM} &=& \frac{3}{2} ( Y_d^\dagger Y_d - Y_u^\dagger Y_u ) + T - \bigg( \frac{1}{4} g_1^2  + \frac{9}{4} g_2^2 + 8 g_3^2 \bigg) \; ,\nonumber \\ 
\beta_e^{SM} &=& \frac{3}{2} ( Y_e^\dagger Y_e ) + T - \frac{9}{4} ( g_1^2 + g_2^2 ) \; ,
\end{eqnarray}
where 
\begin{equation} \label{eq:appendix:T}
 T = {\rm Tr} ( 3 Y_u^\dagger Y_u + 3 Y_d^\dagger Y_d  + Y_e^\dagger Y_e ) \; .
\end{equation}
The one-loop correction to the Yukawa coupling comes from one-loop graphs which contribute to scalar-fermion-fermion vertex.
As is beautifully explained in \cite{Ramond:1999vh}, among these, the contribution $T$ comes from the diagrams involving a fermion loop 
to the scalar propagator while the first set of terms in the above equations come from a Higgs loop correction to the vertex. 
The last set of terms come from the loops involving gauge fields.

In the scenario we are dealing with, we add  $N_f$ vector-like quarks having charge $Q_f$ and the SM singlet S. In model X, we shall be interested in two cases:
(a) $Q_f = 2/3$ and (b) $Q_f > 2/3$. Recall that when $Q_f = 2/3$, the vector-like fermion couples to SM quarks and Higgs with the Yukawa coupling $y_T$ while when 
$Q_f > 2/3$, the Yukawa coupling  $y_T$ is zero.
It is easy to generalise the RG equations presented in \cite{Xiao:2014kba}, we find that
\begin{eqnarray}
\beta_u &=& \beta_u^{SM} + \sum_{i=1}^{N_f} \bigg( \frac{3}{2} {Y_T^{(i)}}^\dagger Y_T^{(i)} + 3 {Y_T^{(i)}}^\dagger Y_T^{(i)} \bigg) \; , \nonumber \\ 
\beta_d &=& \beta_d^{SM}  - \sum_{i=1}^{N_f} \bigg( \frac{3}{2} {Y_T^{(i)}}^\dagger Y_T^{(i)} +  3 {Y_T^{(i)}}^\dagger Y_T^{(i)} \bigg)  \; , \nonumber \\ 
\beta_e &=& \beta_e^{SM} \; . 
\end{eqnarray}
Moreover, the beta functions of the extra Yukawa couplings in Model X are of the form (see also \cite{Zhang:2015uuo})
\begin{eqnarray}
\label{eq:YTyM}
 \beta_{y_T}^{(k)} &=& \sum_{i=1}^{N_f} \frac{9}{2} {Y_T^{(i)}}^\dagger Y_T^{(i)} + \frac{{Y_M^{(k)}}^2}{2} + 
 \frac{3}{2} ( Y_u^\dagger Y_u - Y_d^\dagger Y_d ) \nonumber\\
 && + T - \bigg( \frac{17}{20} g_1^2  + \frac{9}{4} g_2^2 + 8 g_3^2 \bigg) \; , \nonumber \\ 
  \beta_{y_M}^{(k)} &=& \sum_{i=1}^{N_f} 2 \bigg( \frac{3}{2} {Y_M^{(i)}}^2 + 3 {Y_M^{(i)}}^2 \bigg) + {Y_T^{(k)}}^\dagger Y_T^{(k)} \nonumber\\
  && 
 - \bigg( \frac{9}{10} (Y_f)^2 g_1^2 + 8 g_3^2 \bigg) \; .
\end{eqnarray}
where $Y_f$ is the hypercharge of vector-like fermions.  For comparison, one could refer to the appendix of \cite{Xiao:2014kba} to find the corresponding result in the case of a single vector-like fermion.
Similarly, the beta functions of the extra Yukawa couplings in Model Y can be worked out and are of the form 
\begin{eqnarray}
 \beta_{y_F}^{(k)} &=& \sum_{i=1}^{N_f} \frac{9}{2} {Y_F^{(i)}}^\dagger Y_F^{(i)} + \frac{{Y_M^{(k)}}^2}{2} -
 \frac{3}{2} (Y_d^\dagger Y_d ) \nonumber\\
 && + {\rm Tr} ( Y_u^\dagger Y_u + 3 Y_d^\dagger Y_d  + Y_e^\dagger Y_e ) \nonumber\\
 && - \bigg( \frac{17}{20} g_1^2  + \frac{9}{4} g_2^2 + 8 g_3^2 \bigg) \; , \nonumber \\ 
  \beta_{y_M}^{(k)} &=& \sum_{i=1}^{N_f} 2 \bigg( \frac{3}{2} {Y_M^{(i)}}^2 + 4 {Y_M^{(i)}}^2 \bigg) + {Y_F^{(k)}}^\dagger Y_F^{(k)} \nonumber\\
  && 
 - \bigg( \frac{9}{10} (Y_f)^2 g_1^2 +\frac{9}{2} g_2^2 + 8 g_3^2 \bigg) \; ,
\end{eqnarray}

\subsubsection{Modified RGEs for scalar couplings}

The scalar beta functions are related to RG equations by 
\begin{equation}
  \frac{d \lambda}{d t} = \frac{1}{(4 \pi)^2} \beta_\lambda \; .
\end{equation}
In SM, the beta function of the Higgs self-coupling is given by (see, eq (6.48) of \cite{Ramond:1999vh}) 
\begin{eqnarray}
 \beta_\lambda^{SM} & = &  12 \lambda^2 - \bigg( \frac{9}{5} g_1^2 + 9 g_2^2 \bigg) \lambda 
 + \frac{9}{4} \bigg( \frac{3}{25} g_1^4 + \frac{2}{5} g_1^2 g_2^2 + g_2^4 \bigg) \nonumber\\
 && + 4 T \lambda - 4 H \; ,
\end{eqnarray}
where $T$ is as defined in eq (\ref{eq:appendix:T}) and $H$ is given by
\begin{equation}
 H = {\rm Tr} \bigg( 3 (Y_u^\dagger Y_u)^2 + 3 (Y_d^\dagger Y_d)^2  + (Y_e^\dagger Y_e)^2 \bigg) \; .
\end{equation}
In model X, the couplings $\lambda_H$, $\lambda_S$ and $\lambda_{SH}$ evolve in accordance with (see \cite{Xiao:2014kba}) 
\begin{eqnarray}
 \frac{d \lambda_H}{d t} &=& \frac{2}{(4 \pi)^2} \bigg(  
12 \lambda_H^2 + 6 y_t^2 \lambda_H - \frac{9}{10} g_1^2 \lambda_H - \frac{9}{2} g_2^2 \lambda_H \nonumber\\
&& +  
\frac{27}{400} g_1^4 + \frac{9}{16} g_2^4 + \frac{9}{40} g_1^2 g_2^2 - 3 y_t^4  + \frac{\lambda_{SH}^2}{4} \nonumber\\
&& + \sum_{i=1}^{N_f} 6 \lambda_H {y_T^{(i)}}^2    + 
 \sum_{i,j=1}^{N_f} (-3) {y_T^{(i)}}^2 {y_T^{(j)}}^2 \nonumber\\
 && + \sum_{i=1}^{N_f} (-6) {y_T^{(i)}}^2 y_t^2 \bigg) \; , \nonumber \\
 \frac{d \lambda_{SH}}{dt} &=& \frac{2}{(4 \pi)^2} \bigg( \lambda_{SH} 
 \Big( 2 \lambda_{SH} + 6\lambda_H + 3 y_t^2 - \frac{9}{20} g_1^2 - \frac{9}{4} g_2^2 \nonumber \\
 && + 3 \lambda_S + \sum_{i=1}^{N_f} 6 {y_M^{(i)}}^2 + \sum_{i=1}^{N_f} 3 {y_T^{(i)}}^2 \Big) \nonumber\\
 && +  
  \sum_{i=1}^{N_f} (- 12) {y_T^{(i)}}^2 {y_M^{(i)}}^2 \bigg)  \; , \nonumber \\
 \frac{d \lambda_{S}}{dt} &=& \frac{2}{(4 \pi)^2} \bigg( 9 \lambda_S^2 + \lambda_{SH}^2  + \sum_{i=1}^{N_f} 12 \lambda_S {y_M^{(i)}}^2 
 \nonumber\\
 && + \sum_{i=1}^{N_f} (-12) {y_M^{(i)}}^4  \bigg) \; . 
\end{eqnarray}
In model Y, the couplings $\lambda_H$, $\lambda_S$ and $\lambda_{SH}$ evolve in accordance with  
\begin{eqnarray}
 \frac{d \lambda_H}{d t} &=& \frac{2}{(4 \pi)^2} \bigg(  
12 \lambda_H^2 + 6 y_t^2 \lambda_H - \frac{9}{10} g_1^2 \lambda_H - \frac{9}{2} g_2^2 \lambda_H \nonumber\\
&& +  
\frac{27}{400} g_1^4 + \frac{9}{16} g_2^4 + \frac{9}{40} g_1^2 g_2^2 - 3 y_t^4  + \frac{\lambda_{SH}^2}{4} \nonumber\\
&& + \sum_{i=1}^{N_f} 6 \lambda_H {y_F^{(i)}}^2    + 
 \sum_{i,j=1}^{N_f} (-3) {y_F^{(i)}}^2 {y_F^{(j)}}^2 \nonumber\\
 && + \sum_{i=1}^{N_f} (-6) {y_F^{(i)}}^2 y_b^2 \bigg) \; , \nonumber \\
 \frac{d \lambda_{SH}}{dt} &=& \frac{2}{(4 \pi)^2} \bigg( \lambda_{SH} 
 \Big( 2 \lambda_{SH} + 6\lambda_H + 3 y_t^2 - \frac{9}{20} g_1^2 - \frac{9}{4} g_2^2 \nonumber \\
 && + 3 \lambda_S + \sum_{i=1}^{N_f} 6 {y_M^{(i)}}^2 + \sum_{i=1}^{N_f} 3 {y_F^{(i)}}^2 \Big) \nonumber\\
 && +  
  \sum_{i=1}^{N_f} (- 12) {y_F^{(i)}}^2 {y_M^{(i)}}^2 \bigg)  \; , \nonumber \\
 \frac{d \lambda_{S}}{dt} &=& \frac{2}{(4 \pi)^2} \bigg( 9 \lambda_S^2 + \lambda_{SH}^2  + \sum_{i=1}^{N_f} 12 \lambda_S \ {y_M^{(i)}}^2 
 \nonumber\\
 && + \sum_{i=1}^{N_f} (-12) {y_M^{(i)}}^4  \bigg) \; . 
\end{eqnarray}

\acknowledgments

The authors would like to thank S. Mohanty and N. Mahajan (PRL, Ahmedabad) for discussions. The authors would also like to thank anonymous referee for useful comments and suggestions.




\begin{thebibliography}{99}



\bibitem{ATLAS} ATLAS Collaboration, Search for resonances decaying to photon pairs in 3.2 fb$^{-1}$ of pp collisions at
sqrt(s) = 13 TeV with the ATLAS detector," %
 ATLAS-CONF-2015-081 (2015) .

\bibitem{CMS:2015dxe} 
  CMS Collaboration [CMS Collaboration],
  ``Search for new physics in high mass diphoton events in proton-proton collisions at 13TeV,''
  CMS-PAS-EXO-15-004.
  
    
\bibitem{CMS:2015cwa} 
  CMS Collaboration [CMS Collaboration],
  ``Search for High-Mass Diphoton Resonances in pp Collisions at sqrt(s)=8 TeV with the CMS Detector,''
  CMS-PAS-EXO-12-045.
  
\bibitem{Aad:2015mna} 
  G.~Aad {\it et al.} [ATLAS Collaboration],
  ``Search for high-mass diphoton resonances in $pp$ collisions at $\sqrt{s}=8$ TeV with the ATLAS detector,''
  Phys.\ Rev.\ D {\bf 92}, no. 3, 032004 (2015)
  doi:10.1103/PhysRevD.92.032004
  [arXiv:1504.05511 [hep-ex]].
  
\bibitem{DiChiara:2015vdm} 
  S.~Di Chiara, L.~Marzola and M.~Raidal,
  ``First interpretation of the 750 GeV di-photon resonance at the LHC,''
  arXiv:1512.04939 [hep-ph].
  
\bibitem{Landau:1948kw} 
  L.~D.~Landau,
  ``On the angular momentum of a system of two photons,''
  Dokl.\ Akad.\ Nauk Ser.\ Fiz.\  {\bf 60}, no. 2, 207 (1948).
  doi:10.1016/B978-0-08-010586-4.50070-5

\bibitem{Yang:1950rg} 
  C.~N.~Yang,
  ``Selection Rules for the Dematerialization of a Particle Into Two Photons,''
  Phys.\ Rev.\  {\bf 77}, 242 (1950).
  doi:10.1103/PhysRev.77.242
  
\bibitem{Mambrini:2015wyu} 
  Y.~Mambrini, G.~Arcadi and A.~Djouadi,
  ``The LHC diphoton resonance and dark matter,''
  arXiv:1512.04913 [hep-ph].
  
\bibitem{Harigaya:2015ezk} 
  K.~Harigaya and Y.~Nomura,
  ``Composite Models for the 750 GeV Diphoton Excess,''
  arXiv:1512.04850 [hep-ph].
  
\bibitem{Nakai:2015ptz} 
  Y.~Nakai, R.~Sato and K.~Tobioka,
  ``Footprints of New Strong Dynamics via Anomaly,''
  arXiv:1512.04924 [hep-ph].
  
\bibitem{Buttazzo:2015txu} 
  D.~Buttazzo, A.~Greljo and D.~Marzocca,
  ``Knocking on New Physics' door with a Scalar Resonance,''
  arXiv:1512.04929 [hep-ph].
  
\bibitem{Franceschini:2015kwy} 
  R.~Franceschini {\it et al.},
  ``What is the gamma gamma resonance at 750 GeV?,''
  arXiv:1512.04933 [hep-ph].
    
  
\bibitem{Angelescu:2015uiz} 
  A.~Angelescu, A.~Djouadi and G.~Moreau,
  ``Scenarii for interpretations of the LHC diphoton excess: two Higgs doublets and vector-like quarks and leptons,''
  arXiv:1512.04921 [hep-ph].
  
  
\bibitem{Pilaftsis:2015ycr} 
  A.~Pilaftsis,
  ``Diphoton Signatures from Heavy Axion Decays at LHC,''
  arXiv:1512.04931 [hep-ph].
  
\bibitem{Bellazzini:2015nxw} 
  B.~Bellazzini, R.~Franceschini, F.~Sala and J.~Serra,
  ``Goldstones in Diphotons,''
  arXiv:1512.05330 [hep-ph].
  
  
  
\bibitem{Knapen:2015dap} 
  S.~Knapen, T.~Melia, M.~Papucci and K.~Zurek,
  ``Rays of light from the LHC,''
  arXiv:1512.04928 [hep-ph].
  
\bibitem{Ellis:2015oso} 
  J.~Ellis, S.~A.~R.~Ellis, J.~Quevillon, V.~Sanz and T.~You,
  ``On the Interpretation of a Possible $\sim 750$ GeV Particle Decaying into $\gamma \gamma$,''
  arXiv:1512.05327 [hep-ph].
  
\bibitem{Gupta:2015zzs} 
  R.~S.~Gupta, S.~Jäger, Y.~Kats, G.~Perez and E.~Stamou,
  ``Interpreting a 750 GeV Diphoton Resonance,''
  arXiv:1512.05332 [hep-ph].
  
\bibitem{Molinaro:2015cwg} 
  E.~Molinaro, F.~Sannino and N.~Vignaroli,
  ``Minimal Composite Dynamics versus Axion Origin of the Diphoton excess,''
  arXiv:1512.05334 [hep-ph].
  
\bibitem{Higaki:2015jag} 
  T.~Higaki, K.~S.~Jeong, N.~Kitajima and F.~Takahashi,
  ``The QCD Axion from Aligned Axions and Diphoton Excess,''
  arXiv:1512.05295 [hep-ph].
  
\bibitem{McDermott:2015sck} 
  S.~D.~McDermott, P.~Meade and H.~Ramani,
  ``Singlet Scalar Resonances and the Diphoton Excess,''
  arXiv:1512.05326 [hep-ph].
  
\bibitem{Low:2015qep} 
  M.~Low, A.~Tesi and L.~T.~Wang,
  ``A pseudoscalar decaying to photon pairs in the early LHC run 2 data,''
  arXiv:1512.05328 [hep-ph].
  
\bibitem{Petersson:2015mkr} 
  C.~Petersson and R.~Torre,
  ``The 750 GeV diphoton excess from the goldstino superpartner,''
  arXiv:1512.05333 [hep-ph].
  
\bibitem{Cao:2015pto} 
  Q.~H.~Cao, Y.~Liu, K.~P.~Xie, B.~Yan and D.~M.~Zhang,
  ``A Boost Test of Anomalous Diphoton Resonance at the LHC,''
  arXiv:1512.05542 [hep-ph].
  
\bibitem{Matsuzaki:2015che} 
  S.~Matsuzaki and K.~Yamawaki,
  ``750 GeV Diphoton Signal from One-Family Walking Technipion,''
  arXiv:1512.05564 [hep-ph].
  
\bibitem{Dutta:2015wqh} 
  B.~Dutta, Y.~Gao, T.~Ghosh, I.~Gogoladze and T.~Li,
  ``Interpretation of the diphoton excess at CMS and ATLAS,''
  arXiv:1512.05439 [hep-ph].
  
\bibitem{Kobakhidze:2015ldh} 
  A.~Kobakhidze, F.~Wang, L.~Wu, J.~M.~Yang and M.~Zhang,
  ``LHC 750 GeV diphoton resonance explained as a heavy scalar in top-seesaw model,''
  arXiv:1512.05585 [hep-ph].
  
\bibitem{Cox:2015ckc} 
  P.~Cox, A.~D.~Medina, T.~S.~Ray and A.~Spray,
  ``Diphoton Excess at 750 GeV from a Radion in the Bulk-Higgs Scenario,''
  arXiv:1512.05618 [hep-ph].
  
\bibitem{Ahmed:2015uqt} 
  A.~Ahmed, B.~M.~Dillon, B.~Grzadkowski, J.~F.~Gunion and Y.~Jiang,
  ``Higgs-radion interpretation of 750 GeV di-photon excess at the LHC,''
  arXiv:1512.05771 [hep-ph].
  
  
\bibitem{Agrawal:2015dbf} 
  P.~Agrawal, J.~Fan, B.~Heidenreich, M.~Reece and M.~Strassler,
  ``Experimental Considerations Motivated by the Diphoton Excess at the LHC,''
  arXiv:1512.05775 [hep-ph].
  
\bibitem{Becirevic:2015fmu} 
  D.~Becirevic, E.~Bertuzzo, O.~Sumensari and R.~Z.~Funchal,
  ``Can the new resonance at LHC be a CP-Odd Higgs boson?,''
  arXiv:1512.05623 [hep-ph].
  
\bibitem{No:2015bsn} 
  J.~M.~No, V.~Sanz and J.~Setford,
  ``See-Saw Composite Higgses at the LHC: Linking Naturalness to the $750$ GeV Di-Photon Resonance,''
  arXiv:1512.05700 [hep-ph].
  
\bibitem{Demidov:2015zqn} 
  S.~V.~Demidov and D.~S.~Gorbunov,
  ``On sgoldstino interpretation of the diphoton excess,''
  arXiv:1512.05723 [hep-ph].
  
\bibitem{Chao:2015ttq} 
  W.~Chao, R.~Huo and J.~H.~Yu,
  ``The Minimal Scalar-Stealth Top Interpretation of the Diphoton Excess,''
  arXiv:1512.05738 [hep-ph].
  
\bibitem{Fichet:2015vvy} 
  S.~Fichet, G.~von Gersdorff and C.~Royon,
  ``Scattering Light by Light at 750 GeV at the LHC,''
  arXiv:1512.05751 [hep-ph].
  
\bibitem{Curtin:2015jcv} 
  D.~Curtin and C.~B.~Verhaaren,
  ``Quirky Explanations for the Diphoton Excess,''
  arXiv:1512.05753 [hep-ph].
  
\bibitem{Bian:2015kjt} 
  L.~Bian, N.~Chen, D.~Liu and J.~Shu,
  ``A hidden confining world on the 750 GeV diphoton excess,''
  arXiv:1512.05759 [hep-ph].
  
\bibitem{Chakrabortty:2015hff} 
  J.~Chakrabortty, A.~Choudhury, P.~Ghosh, S.~Mondal and T.~Srivastava,
  ``Di-photon resonance around 750 GeV: shedding light on the theory underneath,''
  arXiv:1512.05767 [hep-ph].
  
\bibitem{Csaki:2015vek} 
  C.~Csaki, J.~Hubisz and J.~Terning,
  ``The Minimal Model of a Diphoton Resonance: Production without Gluon Couplings,''
  arXiv:1512.05776 [hep-ph].
  
\bibitem{Falkowski:2015swt} 
  A.~Falkowski, O.~Slone and T.~Volansky,
  ``Phenomenology of a 750 GeV Singlet,''
  arXiv:1512.05777 [hep-ph].
  
\bibitem{Aloni:2015mxa} 
  D.~Aloni, K.~Blum, A.~Dery, A.~Efrati and Y.~Nir,
  ``On a possible large width 750 GeV diphoton resonance at ATLAS and CMS,''
  arXiv:1512.05778 [hep-ph].
  
\bibitem{Bai:2015nbs} 
  Y.~Bai, J.~Berger and R.~Lu,
  ``A 750 GeV Dark Pion: Cousin of a Dark G-parity-odd WIMP,''
  arXiv:1512.05779 [hep-ph].
  
\bibitem{Benbrik:2015fyz} 
  R.~Benbrik, C.~H.~Chen and T.~Nomura,
  ``Higgs singlet as a diphoton resonance in a vector-like quark model,''
  arXiv:1512.06028 [hep-ph].
  
\bibitem{Kim:2015ron} 
  J.~S.~Kim, J.~Reuter, K.~Rolbiecki and R.~R.~de Austri,
  ``A resonance without resonance: scrutinizing the diphoton excess at 750 GeV,''
  arXiv:1512.06083 [hep-ph].
  
\bibitem{Gabrielli:2015dhk} 
  E.~Gabrielli, K.~Kannike, B.~Mele, M.~Raidal, C.~Spethmann and H.~Veermäe,
  ``A SUSY Inspired Simplified Model for the 750 GeV Diphoton Excess,''
  arXiv:1512.05961 [hep-ph].
  
\bibitem{Alves:2015jgx} 
  A.~Alves, A.~G.~Dias and K.~Sinha,
  ``The 750 GeV $S$-cion: Where else should we look for it?,''
  arXiv:1512.06091 [hep-ph].
  
\bibitem{Megias:2015ory} 
  E.~Megias, O.~Pujolas and M.~Quiros,
  ``On dilatons and the LHC diphoton excess,''
  arXiv:1512.06106 [hep-ph].
  
\bibitem{Carpenter:2015ucu} 
  L.~M.~Carpenter, R.~Colburn and J.~Goodman,
  ``Supersoft SUSY Models and the 750 GeV Diphoton Excess, Beyond Effective Operators,''
  arXiv:1512.06107 [hep-ph].
  
\bibitem{Bernon:2015abk} 
  J.~Bernon and C.~Smith,
  ``Could the width of the diphoton anomaly signal a three-body decay ?,''
  arXiv:1512.06113 [hep-ph].
  
  
  
\bibitem{Ringwald:2015dsf} 
  A.~Ringwald and K.~Saikawa,
  ``Accion dark matter in the post-inflationary Peccei-Quinn symmetry breaking scenario,''
  arXiv:1512.06436 [hep-ph].
  
\bibitem{Arun:2015ubr} 
  M.~T.~Arun and P.~Saha,
  ``Gravitons in multiply warped scenarios - at 750 GeV and beyond,''
  arXiv:1512.06335 [hep-ph].
  
\bibitem{Han:2015cty} 
  C.~Han, H.~M.~Lee, M.~Park and V.~Sanz,
  ``The diphoton resonance as a gravity mediator of dark matter,''
  arXiv:1512.06376 [hep-ph].
  
\bibitem{Chang:2015bzc} 
  S.~Chang,
  ``A Simple $U(1)$ Gauge Theory Explanation of the Diphoton Excess,''
  arXiv:1512.06426 [hep-ph].
  
\bibitem{Han:2015dlp} 
  H.~Han, S.~Wang and S.~Zheng,
  ``Scalar Explanation of Diphoton Excess at LHC,''
  arXiv:1512.06562 [hep-ph].
  
\bibitem{Luo:2015yio} 
  M.~x.~Luo, K.~Wang, T.~Xu, L.~Zhang and G.~Zhu,
  ``Squarkonium/Diquarkonium and the Di-photon Excess,''
  arXiv:1512.06670 [hep-ph].
  
\bibitem{Chang:2015sdy} 
  J.~Chang, K.~Cheung and C.~T.~Lu,
  ``Interpreting the 750 GeV Di-photon Resonance using photon-jets in Hidden-Valley-like models,''
  arXiv:1512.06671 [hep-ph].
  
\bibitem{Bardhan:2015hcr} 
  D.~Bardhan, D.~Bhatia, A.~Chakraborty, U.~Maitra, S.~Raychaudhuri and T.~Samui,
  ``Radion Candidate for the LHC Diphoton Resonance,''
  arXiv:1512.06674 [hep-ph].
  
\bibitem{Feng:2015wil} 
  T.~F.~Feng, X.~Q.~Li, H.~B.~Zhang and S.~M.~Zhao,
  ``The LHC 750 GeV diphoton excess in supersymmetry with gauged baryon and lepton numbers,''
  arXiv:1512.06696 [hep-ph].
  
\bibitem{Barducci:2015gtd} 
  D.~Barducci, A.~Goudelis, S.~Kulkarni and D.~Sengupta,
  ``One jet to rule them all: monojet constraints and invisible decays of a 750 GeV diphoton resonance,''
  arXiv:1512.06842 [hep-ph].
  
\bibitem{Chao:2015nsm} 
  W.~Chao,
  ``Symmetries Behind the 750 GeV Diphoton Excess,''
  arXiv:1512.06297 [hep-ph].
  
\bibitem{Chakraborty:2015jvs} 
  I.~Chakraborty and A.~Kundu,
  ``Diphoton excess at 750 GeV: Singlet scalars confront naturalness,''
  arXiv:1512.06508 [hep-ph].
  
\bibitem{Ding:2015rxx} 
  R.~Ding, L.~Huang, T.~Li and B.~Zhu,
  ``Interpreting $750$ GeV Diphoton Excess with R-parity Violation Supersymmetry,''
  arXiv:1512.06560 [hep-ph].
  
\bibitem{Hatanaka:2015qjo} 
  H.~Hatanaka,
  ``Oblique corrections from less-Higgsless models in warped space,''
  arXiv:1512.06595 [hep-ph].
  
\bibitem{Antipin:2015kgh} 
  O.~Antipin, M.~Mojaza and F.~Sannino,
  ``A natural Coleman-Weinberg theory explains the diphoton excess,''
  arXiv:1512.06708 [hep-ph].
  
\bibitem{Wang:2015kuj} 
  F.~Wang, L.~Wu, J.~M.~Yang and M.~Zhang,
  ``750 GeV Diphoton Resonance, 125 GeV Higgs and Muon g-2 Anomaly in Deflected Anomaly Mediation SUSY Breaking Scenario,''
  arXiv:1512.06715 [hep-ph].
  
  
\bibitem{Cao:2015twy} 
  J.~Cao, C.~Han, L.~Shang, W.~Su, J.~M.~Yang and Y.~Zhang,
  ``Interpreting the 750 GeV diphoton excess by the singlet extension of the Manohar-Wise Model,''
  arXiv:1512.06728 [hep-ph].
  
\bibitem{Huang:2015evq} 
  F.~P.~Huang, C.~S.~Li, Z.~L.~Liu and Y.~Wang,
  ``750 GeV Diphoton Excess from Cascade Decay,''
  arXiv:1512.06732 [hep-ph].
  
\bibitem{Bi:2015uqd} 
  X.~J.~Bi, Q.~F.~Xiang, P.~F.~Yin and Z.~H.~Yu,
  ``The 750 GeV diphoton excess at the LHC and dark matter constraints,''
  arXiv:1512.06787 [hep-ph].
  
\bibitem{Kim:2015ksf} 
  J.~S.~Kim, K.~Rolbiecki and R.~R.~de Austri,
  ``Model-independent combination of diphoton constraints at 750 GeV,''
  arXiv:1512.06797 [hep-ph].
  
\bibitem{Berthier:2015vbb} 
  L.~Berthier, J.~M.~Cline, W.~Shepherd and M.~Trott,
  ``Effective interpretations of a diphoton excess,''
  arXiv:1512.06799 [hep-ph].
  
\bibitem{Cline:2015msi} 
  J.~M.~Cline and Z.~Liu,
  ``LHC diphotons from electroweakly pair-produced composite pseudoscalars,''
  arXiv:1512.06827 [hep-ph].
  
\bibitem{Chala:2015cev} 
  M.~Chala, M.~Duerr, F.~Kahlhoefer and K.~Schmidt-Hoberg,
  ``Tricking Landau-Yang: How to obtain the diphoton excess from a vector resonance,''
  arXiv:1512.06833 [hep-ph].
  
\bibitem{Kulkarni:2015gzu} 
  K.~Kulkarni,
  ``Extension of $\nu$MSM model and possible explanations of recent astronomical and collider observations,''
  arXiv:1512.06836 [hep-ph].
  
\bibitem{Dev:2015isx} 
  P.~S.~B.~Dev and D.~Teresi,
  ``Asymmetric Dark Matter in the Sun and the Diphoton Excess at the LHC,''
  arXiv:1512.07243 [hep-ph].
  
\bibitem{Boucenna:2015pav} 
  S.~M.~Boucenna, S.~Morisi and A.~Vicente,
  ``The LHC diphoton resonance from gauge symmetry,''
  arXiv:1512.06878 [hep-ph].
  
\bibitem{deBlas:2015hlv} 
  J.~de Blas, J.~Santiago and R.~Vega-Morales,
  ``New vector bosons and the diphoton excess,''
  arXiv:1512.07229 [hep-ph].
  
\bibitem{Murphy:2015kag} 
  C.~W.~Murphy,
  ``Vector Leptoquarks and the 750 GeV Diphoton Resonance at the LHC,''
  arXiv:1512.06976 [hep-ph].
  
\bibitem{Hernandez:2015ywg} 
  A.~E.~C.~Hernández and I.~Nisandzic,
  ``LHC diphoton 750 GeV resonance as an indication of $SU(3)_c\times SU(3)_L\times U(1)_X$ gauge symmetry,''
  arXiv:1512.07165 [hep-ph].
  
\bibitem{Dey:2015bur} 
  U.~K.~Dey, S.~Mohanty and G.~Tomar,
  ``750 GeV resonance in the Dark Left-Right Model,''
  arXiv:1512.07212 [hep-ph].
  
\bibitem{Pelaggi:2015knk} 
  G.~M.~Pelaggi, A.~Strumia and E.~Vigiani,
  ``Trinification can explain the di-photon and di-boson LHC anomalies,''
  arXiv:1512.07225 [hep-ph].
  
\bibitem{Belyaev:2015hgo} 
  A.~Belyaev, G.~Cacciapaglia, H.~Cai, T.~Flacke, A.~Parolini and H.~Serôdio,
  ``Singlets in Composite Higgs Models in light of the LHC di-photon searches,''
  arXiv:1512.07242 [hep-ph].
  
\bibitem{Huang:2015rkj} 
  W.~C.~Huang, Y.~L.~S.~Tsai and T.~C.~Yuan,
  ``Gauged Two Higgs Doublet Model confronts the LHC 750 GeV di-photon anomaly,''
  arXiv:1512.07268 [hep-ph].
  
  
\bibitem{Cao:2015xjz} 
  Q.~H.~Cao, S.~L.~Chen and P.~H.~Gu,
  ``Strong CP Problem, Neutrino Masses and the 750 GeV Diphoton Resonance,''
  arXiv:1512.07541 [hep-ph].
  
\bibitem{Gu:2015lxj} 
  J.~Gu and Z.~Liu,
  ``Running after Diphoton,''
  arXiv:1512.07624 [hep-ph].
  
\bibitem{Moretti:2015pbj} 
  S.~Moretti and K.~Yagyu,
  ``The 750 GeV diphoton excess and its explanation in 2-Higgs Doublet Models with a real inert scalar multiplet,''
  arXiv:1512.07462 [hep-ph].
  
\bibitem{Patel:2015ulo} 
  K.~M.~Patel and P.~Sharma,
  ``Interpreting 750 GeV diphoton excess in SU(5) grand unified theory,''
  arXiv:1512.07468 [hep-ph].
  
\bibitem{Badziak:2015zez} 
  M.~Badziak,
  ``Interpreting the 750 GeV diphoton excess in minimal extensions of Two-Higgs-Doublet models,''
  arXiv:1512.07497 [hep-ph].
  
\bibitem{Chakraborty:2015gyj} 
  S.~Chakraborty, A.~Chakraborty and S.~Raychaudhuri,
  ``Diphoton resonance at 750 GeV in the broken MRSSM,''
  arXiv:1512.07527 [hep-ph].
  
  
\bibitem{Altmannshofer:2015xfo} 
  W.~Altmannshofer, J.~Galloway, S.~Gori, A.~L.~Kagan, A.~Martin and J.~Zupan,
  ``On the 750 GeV di-photon excess,''
  arXiv:1512.07616 [hep-ph].
  
\bibitem{Cvetic:2015vit} 
  M.~Cveti?, J.~Halverson and P.~Langacker,
  ``String Consistency, Heavy Exotics, and the $750$ GeV Diphoton Excess at the LHC,''
  arXiv:1512.07622 [hep-ph].
  
\bibitem{Allanach:2015ixl} 
  B.~C.~Allanach, P.~S.~B.~Dev, S.~A.~Renner and K.~Sakurai,
  ``Di-photon Excess Explained by a Resonant Sneutrino in R-parity Violating Supersymmetry,''
  arXiv:1512.07645 [hep-ph].
  
  
\bibitem{Davoudiasl:2015cuo} 
  H.~Davoudiasl and C.~Zhang,
  ``A 750 GeV Messenger of Dark Conformal Symmetry Breaking,''
  arXiv:1512.07672 [hep-ph].
  
\bibitem{Das:2015enc} 
  K.~Das and S.~K.~Rai,
  ``The 750 GeV Diphoton excess in a $U(1)$ hidden symmetry model,''
  arXiv:1512.07789 [hep-ph].
  
\bibitem{Cheung:2015cug} 
  K.~Cheung, P.~Ko, J.~S.~Lee, J.~Park and P.~Y.~Tseng,
  ``A Higgcision study on the 750 GeV Di-photon Resonance and 125 GeV SM Higgs boson with the Higgs-Singlet Mixing,''
  arXiv:1512.07853 [hep-ph].
  
\bibitem{Liu:2015yec} 
  J.~Liu, X.~P.~Wang and W.~Xue,
  ``LHC diphoton excess from colorful resonances,''
  arXiv:1512.07885 [hep-ph].
  
  
\bibitem{Zhang:2015uuo} 
  J.~Zhang and S.~Zhou,
  ``Electroweak Vacuum Stability and Diphoton Excess at 750 GeV,''
  arXiv:1512.07889 [hep-ph].
  
\bibitem{Casas:2015blx} 
  J.~A.~Casas, J.~R.~Espinosa and J.~M.~Moreno,
  ``The 750 GeV Diphoton Excess as a First Light on Supersymmetry Breaking,''
  arXiv:1512.07895 [hep-ph].
  
\bibitem{Hall:2015xds} 
  L.~J.~Hall, K.~Harigaya and Y.~Nomura,
  ``750 GeV Diphotons: Implications for Supersymmetric Unification,''
  arXiv:1512.07904 [hep-ph].
  
    
\bibitem{Dijkstra:2004cc} 
  T.~P.~T.~Dijkstra, L.~R.~Huiszoon and A.~N.~Schellekens,
  ``Supersymmetric standard model spectra from RCFT orientifolds,''
  Nucl.\ Phys.\ B {\bf 710}, 3 (2005)
  doi:10.1016/j.nuclphysb.2004.12.032
  [hep-th/0411129].
  
\bibitem{Lebedev:2006kn} 
  O.~Lebedev, H.~P.~Nilles, S.~Raby, S.~Ramos-Sanchez, M.~Ratz, P.~K.~S.~Vaudrevange and A.~Wingerter,
  ``A Mini-landscape of exact MSSM spectra in heterotic orbifolds,''
  Phys.\ Lett.\ B {\bf 645}, 88 (2007)
  doi:10.1016/j.physletb.2006.12.012
  [hep-th/0611095].
   
\bibitem{Han:2003wu} 
  T.~Han, H.~E.~Logan, B.~McElrath and L.~T.~Wang,
  ``Phenomenology of the little Higgs model,''
  Phys.\ Rev.\ D {\bf 67}, 095004 (2003)
  doi:10.1103/PhysRevD.67.095004
  [hep-ph/0301040].
  
\bibitem{Contino:2006qr} 
  R.~Contino, L.~Da Rold and A.~Pomarol,
  ``Light custodians in natural composite Higgs models,''
  Phys.\ Rev.\ D {\bf 75}, 055014 (2007)
  doi:10.1103/PhysRevD.75.055014
  [hep-ph/0612048].
  
    
\bibitem{CMS:2015alb} 
  CMS Collaboration [CMS Collaboration],
  ``Search for top quark partners with charge 5/3 at $\sqrt{s}=13$ TeV,''
  CMS-PAS-B2G-15-006.
  
\bibitem{Khachatryan:2015oba} 
  V.~Khachatryan {\it et al.} [CMS Collaboration],
  ``Search for Vector-Like Charge 2/3 T Quarks in Proton-Proton Collisions at $\sqrt{s}$ = 8 TeV,''
  arXiv:1509.04177 [hep-ex].
  
\bibitem{Aad:2016qpo} 
  G.~Aad {\it et al.} [ATLAS Collaboration],
  arXiv:1602.05606 [hep-ex].
  
\bibitem{Gunion:1989we} 
  J.~F.~Gunion, H.~E.~Haber, G.~L.~Kane and S.~Dawson,
  ``The Higgs Hunter's Guide,''
  Front.\ Phys.\  {\bf 80}, 1 (2000).
  
  
      

    
    
  
  

  
\bibitem{Xiao:2014kba} 
  M.~L.~Xiao and J.~H.~Yu,
  ``Stabilizing electroweak vacuum in a vectorlike fermion model,''
  Phys.\ Rev.\ D {\bf 90}, no. 1, 014007 (2014)
  [Phys.\ Rev.\ D {\bf 90}, no. 1, 019901 (2014)]
  doi:10.1103/PhysRevD.90.014007, 10.1103/PhysRevD.90.019901
  [arXiv:1404.0681 [hep-ph]].
  
\bibitem{EliasMiro:2012ay} 
  J.~Elias-Miro, J.~R.~Espinosa, G.~F.~Giudice, H.~M.~Lee and A.~Strumia,
  ``Stabilization of the Electroweak Vacuum by a Scalar Threshold Effect,''
  JHEP {\bf 1206}, 031 (2012)
  doi:10.1007/JHEP06(2012)031
  [arXiv:1203.0237 [hep-ph]].
  
  

\bibitem{Hamada:2015skp} 
  Y.~Hamada, T.~Noumi, S.~Sun and G.~Shiu,
  arXiv:1512.08984 [hep-ph].
  

\bibitem{Holthausen:2011aa} 
  M.~Holthausen, K.~S.~Lim and M.~Lindner,
  ``Planck scale Boundary Conditions and the Higgs Mass,''
  JHEP {\bf 1202}, 037 (2012)
  doi:10.1007/JHEP02(2012)037
  [arXiv:1112.2415 [hep-ph]].
  
\bibitem{EliasMiro:2011aa} 
  J.~Elias-Miro, J.~R.~Espinosa, G.~F.~Giudice, G.~Isidori, A.~Riotto and A.~Strumia,
  ``Higgs mass implications on the stability of the electroweak vacuum,''
  Phys.\ Lett.\ B {\bf 709}, 222 (2012)
  doi:10.1016/j.physletb.2012.02.013
  [arXiv:1112.3022 [hep-ph]].


  
\bibitem{Buttazzo:2013uya} 
  D.~Buttazzo, G.~Degrassi, P.~P.~Giardino, G.~F.~Giudice, F.~Sala, A.~Salvio and A.~Strumia,
  ``Investigating the near-criticality of the Higgs boson,''
  JHEP {\bf 1312}, 089 (2013)
  doi:10.1007/JHEP12(2013)089
  [arXiv:1307.3536 [hep-ph]].

\bibitem{Goswami:2014hoa} 
  G.~Goswami and S.~Mohanty,
  ``Higgs instability and de Sitter radiation,''
  Phys.\ Lett.\ B {\bf 751}, 113 (2015)
  doi:10.1016/j.physletb.2015.10.027
  [arXiv:1406.5644 [hep-ph]].


\bibitem{Ade:2015tva} 
  P.~A.~R.~Ade {\it et al.} [BICEP2 and Planck Collaborations],
  ``Joint Analysis of BICEP2/$Keck ?Array$ and $Planck$ Data,''
  Phys.\ Rev.\ Lett.\  {\bf 114}, 101301 (2015)
  doi:10.1103/PhysRevLett.114.101301
  [arXiv:1502.00612 [astro-ph.CO]].

\bibitem{Array:2015xqh} 
  P.~A.~R.~Ade {\it et al.} [BICEP2 and Keck Array Collaborations],
  ``BICEP2 / Keck Array VI: Improved Constraints On Cosmology and Foregrounds When Adding 95 GHz Data From Keck Array,''
  arXiv:1510.09217 [astro-ph.CO].

	
\bibitem{Bezrukov:2014ipa} 
  F.~Bezrukov, J.~Rubio and M.~Shaposhnikov,
  ``Living beyond the edge: Higgs inflation and vacuum metastability,''
  Phys.\ Rev.\ D {\bf 92}, no. 8, 083512 (2015)
  doi:10.1103/PhysRevD.92.083512
  [arXiv:1412.3811 [hep-ph]].
  
\bibitem{Bezrukov:2013fka} 
  F.~Bezrukov,
  Class.\ Quant.\ Grav.\  {\bf 30}, 214001 (2013)
  doi:10.1088/0264-9381/30/21/214001
  [arXiv:1307.0708 [hep-ph]].

\bibitem{Salvio:2015kka} 
  A.~Salvio and A.~Mazumdar,
  ``Classical and Quantum Initial Conditions for Higgs Inflation,''
  Phys.\ Lett.\ B {\bf 750}, 194 (2015)
  doi:10.1016/j.physletb.2015.09.020
  [arXiv:1506.07520 [hep-ph]].


\bibitem{Falkowski:2015iwa} 
  A.~Falkowski, C.~Gross and O.~Lebedev,
  ``A second Higgs from the Higgs portal,''
  JHEP {\bf 1505}, 057 (2015)
  doi:10.1007/JHEP05(2015)057
  [arXiv:1502.01361 [hep-ph]].
  
\bibitem{Herranen:2015ima} 
  M.~Herranen, T.~Markkanen, S.~Nurmi and A.~Rajantie,
  Phys.\ Rev.\ Lett.\  {\bf 115}, 241301 (2015)
  doi:10.1103/PhysRevLett.115.241301
  [arXiv:1506.04065 [hep-ph]].
  
\bibitem{McDonald:2016cdh} 
  J.~McDonald,
  arXiv:1604.01711 [hep-ph].
  
  
\bibitem{Halyo:1999wq} 
  V.~Halyo, P.~Kim, E.~R.~Lee, I.~T.~Lee, D.~Loomba and M.~L.~Perl,
  Phys.\ Rev.\ Lett.\  {\bf 84}, 2576 (2000)
  doi:10.1103/PhysRevLett.84.2576
  [hep-ex/9910064].
  
\bibitem{Kang:2006yd} 
  J.~Kang, M.~A.~Luty and S.~Nasri,
  JHEP {\bf 0809}, 086 (2008)
  doi:10.1088/1126-6708/2008/09/086
  [hep-ph/0611322].
   
\bibitem{ArkaniHamed:2004fb} 
  N.~Arkani-Hamed and S.~Dimopoulos,
  JHEP {\bf 0506}, 073 (2005)
  doi:10.1088/1126-6708/2005/06/073
  [hep-th/0405159].



  
 
  
\bibitem{Srednicki:2007qs} 
  M.~Srednicki,
  ``Quantum Field Theory,'' Cambridge University Press (2010).

\bibitem{Ramond:1999vh} 
  P.~Ramond,
  "Journeys Beyond The Standard Model (Frontiers in Physics)" {\bf 101} (2003), Westview Press.
  ``Journeys beyond the standard model,''


\end{thebibliography}
\end{document}